\documentclass[aps,prd,twocolumn,superscriptaddress,groupedaddress,nofootinbib]{revtex4}
\usepackage{graphicx}  
\usepackage{dcolumn}   
\usepackage{bm}        
\usepackage{amssymb}   
\usepackage{amsmath}
\usepackage{dsfont}
\usepackage{xcolor}
\usepackage{amsthm}
\usepackage{enumitem}

\usepackage[colorlinks]{hyperref}
\hypersetup{
     breaklinks=true,
    pdfstartview={FitH},    
    colorlinks=true,       
    linkcolor=blue,          
    citecolor=red,        
    filecolor=magenta,      
    urlcolor=blue,           
    anchorcolor=green,      
    linktocpage=true
}

\newcommand{\be}{\begin{equation}}\newcommand{\ee}{\end{equation}}
\newcommand{\bea}{\begin{eqnarray}}\newcommand{\eea}{\end{eqnarray}}
\newcommand{\brr}{\begin{array}}\newcommand{\err}{\end{array}}
\newcommand{\bit}{\begin{itemize}}\newcommand{\eit}{\end{itemize}}
\newcommand{\ben}{\begin{enumerate}}\newcommand{\een}{\end{enumerate}}

\newcommand{\ba}{\begin{array}}
\newcommand{\ea}{\end{array}}

\definecolor{darkred}{rgb}{.8,0,0}
\newcommand{\cdred}{\color{darkred}}

\definecolor{darkblue}{rgb}{0,0,.7}

\def\lf{\left}

\def\ri{\right}
\def\ga{\gamma}
\def\de{\delta}

\def\1{{_{1}}}\def\2{{_{2}}}

\def\noHe0{:\;\!\!\;\!\!:H_e(0):\;\!\!\;\!\!:}
\def\noHm0{:\;\!\!\;\!\!:H_\mu(0):\;\!\!\;\!\!:}

\def\lf{\left}

\def\ri{\right}

\def\ga{\gamma}
\def\de{\delta}

\begin{document}

\title{Coherent states for generalized uncertainty relations as Tsallis probability amplitudes: new route to  non-extensive  thermostatistics}

\author{Petr Jizba}
\email{p.jizba@fjfi.cvut.cz}
\affiliation{FNSPE,
Czech Technical University in Prague, B\v{r}ehov\'{a} 7, 115 19, Prague, Czech Republic}
\affiliation{ITP, Freie Universit\"{a}t Berlin, Arnimallee 14, D-14195 Berlin, Germany}

\author{Gaetano Lambiase}
\email{lambiase@sa.infn.it}
\affiliation{Dipartimento di Fisica, Universit\`a di Salerno, Via Giovanni Paolo II, 132 I-84084 Fisciano (SA), Italy}
\affiliation{INFN, Sezione di Napoli, Gruppo collegato di Salerno, Italy}

\author{Giuseppe Gaetano Luciano}
\email{giuseppegaetano.luciano@udl.cat}
\affiliation{Applied Physics Section of Environmental Science Department, Escola Polit\`ecnica Superior, Universitat de Lleida, Av. Jaume
II, 69, 25001 Lleida, Spain}

\author{Luciano Petruzziello}
\email{lupetruzziello@unisa.it}
\affiliation{INFN, Sezione di Napoli, Gruppo collegato di Salerno, Italy}
\affiliation{Dipartimento di Ingegneria, Universit\`a di Salerno, Via Giovanni Paolo II, 132 I-84084 Fisciano (SA), Italy}
\affiliation{Institut f\"ur Theoretische Physik, Albert-Einstein-Allee 11, Universit\"at Ulm, 89069 Ulm, Germany}

\date{\today}

\begin{abstract}
We study coherent states associated to a generalized uncertainty principle (GUP). We separately analyze the cases of positive and negative deformation parameter $\beta$, showing that the ensuing probability distribution is a Tsallis distribution whose non-extensivity parameter $q$ is monotonically related to $\beta$. Moreover, for $\beta <0$ (corresponding to $q<1$), we reformulate the GUP in terms of a one-parameter class of Tsallis entropy-power based uncertainty relations, which are again saturated by the GUP coherent states.
%
%
We argue that this combination of coherent states with Tsallis entropy offers a natural conceptual framework allowing to  study quasi-classical regime of GUP in terms of non-extensive thermodynamics. We substantiate our claim by discussing generalization of Verlinde's entropic force and ensuing implications in the late-inflation epoch.
%
Corresponding dependence of the  $\beta$ parameter on cosmological time is derived for the reheating epoch. The obtained $\beta$ is consistent with values predicted by both string-theory models and the naturalness principle. 
Further salient issues, including derivation of new $\beta$-dependent expressions for the lowest possible value of the spin and Immirzi parameter in Loop Quantum Gravity, and connection of our proposal with the Magueijo--Smolin doubly
special relativity are also discussed. 
This article provides a more extended and comprehensive treatment of our recent letter [Phys. Rev. D {\bf 105}, L121501 (2022)].
\end{abstract}

\vskip -1.0 truecm

\maketitle

\section{Introduction}
\label{Intro}
The Heisenberg Uncertainty Principle (HUP) -- the cornerstone of quantum mechanics -- provides an intrinsic limitation to the 
simultaneous knowledge of position and momentum  
of any quantum system. While working successfully in low-energy regime, 
it is expected to be modified approaching the Planck scale, due to quantum gravitational effects possibly coming into play. Several models of
quantum gravity, such as String Theory, Loop Quantum Gravity, Quantum
Geometry and Doubly Special Relativity, have converged on the idea
that the HUP should be generalized so as to account
for the emergence of a minimal length at Planck scale. The ensuing uncertainty relations
are typically referred to as Generalized Uncertainty Principles (GUPs).

The simplest version of GUP can be obtained by adding a term quadratic
in the momentum uncertainty over the standard Heisenberg limit~\cite{VenezGrossMende,Gross,Ciafa,Koni,MM,Kempf,FS,Adler2,Capoz,MS1,MS2,Bojo}, i.e.,
\begin{eqnarray}
\label{gup}
\de x\,\de p \, \geq \, \frac{\hbar}{2}\lf(1 \, + \, \beta\,\frac{\de p^2}{m_p^2}\ri)\, ,
\end{eqnarray}
where $m_p =\sqrt{\hbar c/G} \approx 2.2\times 10^{-8}$ Kg is the Planck mass and with $c=1$. 
The (dimensionless) $\beta$ parameter quantifies the
departure from HUP. Note that such term is not fixed by the theory, 
albeit it is generally assumed to be of order unity~\cite{VenezGrossMende,Gross,Ciafa,Koni}.  
However, many theoretical~\cite{Das,Brau,Pedram,ScardCas,Bosso,QC,Scardigli:2018jlm,Petroz,DaRocha,Buoninf,CGgrav,Chung:2019raj,fabian,LucBosFri} and experimental~\cite{Bruk,AliTest,Bawaj,Pendu,GravBar,BossoLigo,LucLuc,Primord}
studies are being developed to 
infer the magnitude of $\beta$, as well as its sign~\cite{Kempf,MS1,Bojo,CGgrav,JKS,KLVY,Ong,illu,illu2}. Clearly, the traditional 
quantum mechanical limit is recovered for $\beta\rightarrow0$ and/or $\delta p\ll m_p$.

The symbol $\delta$ appearing in Eq.~\eqref{gup} denotes 
uncertainty of a given observable and it does not need to be {\em a priori} related to the standard deviation. In fact, in the original Heisenberg relation
$\delta$ can represent Heisenberg's ``ungenauigkeiten'' (i.e., error-disturbance uncertainties caused by the back-reaction in simultaneous measurement of $x$ and $p$) or $\delta p = \langle \psi ||p| | \psi \rangle \equiv \langle |p| \rangle_{\psi} $, see, e.g.~\cite{Maggiore:93}. Nevertheless, in cases when $\delta$ 
is identified with the standard deviation (henceforth denoted by $\Delta$), 
the generalized uncertainty principle~\eqref{gup} directly follows from
the deformed commutation relation (DCR)
\begin{eqnarray}\label{comm}
&&\lf[\hat{x},\hat{p}\ri] \,=\,  i\hbar\lf(1 + \beta\,\frac{\hat{p}^2}{m_p^2}\ri)\, ,
\label{1.cf}
\end{eqnarray}
via the Cauchy--Schwarz inequality~\cite{Schrodinger,Robertson,Muk}, provided one restricts the attention to {\em mirror symmetric} states
satisfying $\langle \hat{p}\rangle_{\psi}=0$. Commutator (\ref{comm}) is typically 
supplemented with the commutators
\begin{eqnarray}
\lf[\hat{x},\hat{x}\ri] \, = \, \lf[\hat{p},\hat{p}\ri] \, = \, 0\, ,
\label{2.cf}
\end{eqnarray}
which, together with (\ref{comm}), satisfy Jacobi identities (up to the lowest-order approximation in $\beta\delta p^2/m^2_p$) and determine the whole symplectic structure of the model.

So far, the quadratic GUP~\eqref{gup} has been largely used to 
study phenomenology of quantum gravity in many sectors, ranging from  
quantum mechanics~\cite{Bernardo,Panella,Ijmpd,bpw} to
particle physics~\cite{Das,Husain,BossoLuciano} and 
cosmology~\cite{JizSc,DFLV,Giardino}. Applications
have also been studied in non-linear optics and condensed matter~\cite{Tawfik,Braidotti,iorio}, where the mass parameter $m_p$
should actually be identified with the effective mass scale in the
quantum description at hand. As an example, we might mention a 1$D$ lattice 
where the usual Weyl-Heisenberg algebra $W_1$ for $\hat p$ and $\hat x$ operators is deformed to the Euclidean algebra $E(2)$, which entails  
GUP with $m_p$ being related to an inverse lattice spacing at small $p$ (momenta deep
inside the Brillouin zone).
Conversely, for large momenta (near the border
of the Brillouin zone), such GUP drifts away from the
quadratic GUP and becomes linear in $p$~\cite{JKS}. 

On the other hand, comparably less attention has been devoted to the
analysis of the quasi-classical domain of the GUP. This
regime has, however, relevant and potentially observable implications 
in the early Universe cosmology and ensuing astrophysics~\cite{Kiefer,Kiefer2}. 
In order to probe physics in the quasi-classical domain, it is customary to rely on 
coherent states (CSs). CSs are, in a sense, privileged quantum states in the description of quantum-to-classical transition, as they are the only states that remain
pure in the decoherence process~\cite{Dutra,Matacz}. 
Since CSs are pure, they allow for maximal resolution in phase-space, 
thus appearing as the closest quantum
counterparts of classical points. Additionally, the CS formalism
offers a convenient description which can draw
upon developments in quantum optics~\cite{Mandel}.

To grasp the core aspects of GUP phenomenology in 
the quasi-classical quantum regime, 
it should be  emphasized that the deformed commutator~\eqref{comm} 
significantly affects the phase-space structure of any quantum system, 
yielding non-trivial implications at microscopic level. 
For instance, in~\cite{Chang} it has been shown that Eq.~\eqref{comm} modifies
the elementary cell volume of each quantum state, which becomes momentum-dependent. This motivates a reformulation of quantum statistical
mechanics when considering GUP.
Preliminary attempts to derive 
the statistics that emerges from the phase-space cell volume implied by the GUP
have been conducted in~\cite{Shababi,LucianoTs}, where the
generalized statistics with a quadratic correction over Gaussian profile emerges quite naturally if one assumes Eq.~\eqref{comm} along with the
condition of invariance of the total phase-space volume.
The question thus arises as to how the interconnection between quantum and statistical properties can be 
rigorously formalized in the quasi-classical regime of GUP. 

In order to tackle the aforementioned issues, we investigate in this work 
possible observational effects of GUP systems in their decoherence domain. 
For this purpose, we first introduce the Schr\"{o}dinger--Nieto-type of minimum-uncertainty  CSs~\cite{Schrodinger,Nieto:78} associated with GUP. 
Subsequently, we show that, in momentum representation, these states
coincide with probability amplitudes derived in Tsallis statistics, which is a non-extensive generalization of Boltzmann--Gibbs statistics based on a non-additive redefinition of the entropy~\cite{Tsallis1,Tsallis2,Tsallis3,jizba-korbel:19,Abe1,JKZ,Caratheodory,CaratheodoryII}. Furthermore, by using the Bekner-Babenko inequality, we recast the GUP for $\beta<0$ in terms of a one-parameter class of Tsallis entropy-power-based uncertainty relations (EPUR), 
which are also saturated by the GUP CSs. By invoking the 
Maximum Entropy Principle (MEP), which states that the 
thermodynamic entropy is
the statistical entropy evaluated at the maximal entropy
distribution, we infer that the combination
of GUP CSs with Tsallis entropy naturally allows us to describe
the quasi-classical domain of GUP in terms of non-extensive Tsallis thermodynamics (NTT). To elucidate our point, 
we discuss three pertinent examples from cosmology:
a) the GUP generalization of Verlinde's entropic gravity force~\cite{Verlinde} and its connection with conformal gravity (CG)~\cite{mannheim,mannheim_a,mannheim_b,mannheim_c,Kazanas} and early Universe cosmology; b) Magueijo--Smolin doubly special relativity (DSR) and finally; c)  we derive 
new $\beta$-dependent expressions for the lowest possible value of the spin and Immirzi parameter
in Loop Quantum Gravity (LQG), discussing their relevance for the gauge group structure of the spin networks in this theory.

The remainder of the work is organized as follows: in Sec.~\ref{CSGUP}
we derive coherent states for the quadratic GUP with both $\beta>0$ and $\beta<0$. A large portion of the section is dedicated to exploring the properties of these CSs. 
In Section~\ref{Tsallis}, we present some fundamentals of non-extensive Tsallis thermodynamics. In particular, we pay a particular attention to an integrating factor for the heat one-form and show that in contrast to conventional thermodynamics it factorizes into thermal and entropic part. The connection
between GUP and Tsallis entropy-power based uncertainty relations 
is investigated in Sec.~\ref{EUR}, while in Sec.~\ref{CosmApp} we discuss some
illustrative examples from the early Universe cosmology.
Conclusions and outlook are finally summarized in Sec.~\ref{Conc}.
For the sake of clarity, we relegate
some more technical considerations to three appendices. 



\section{Coherent states for GUP}
\label{CSGUP}

We shall start with a short outline of the basic  steps that lead from the DCR~(\ref{comm}) to  GUP~(\ref{gup}). To this end, we will assume that  $\delta$ in (\ref{gup}) is identified with the standard deviation computed with respect to some density matrix $\varrho$.

For the variance, (i.e., square of the standard deviation) of an observable $\mathcal{\hat{A}}$ we can write
\begin{eqnarray}\label{var}
(\Delta\mathcal{\hat{A}})^2_{{\varrho}} &\equiv& \mbox{Tr}(\mathcal{\hat{A}}^2\hat{\varrho}) -\mbox{Tr}(\mathcal{\hat{A}}\hat{\varrho})^2\ \nonumber \\[2mm]&=&
\int_{\mathbb{R}} \left(\lambda - \langle\mathcal{\hat{A}}\rangle^2_{\varrho} \right) \ \! d{\mbox{Tr}}\!\left(E_{\lambda}^{(\mathcal{\hat{A}})}\! \hat{\varrho} \right)\, ,
\end{eqnarray}
where $E_{\lambda}^{(\mathcal{\hat{A}})}$ represents the projection–valued measure of $\mathcal{\hat{A}}$ corresponding to spectral value $\lambda$.
If we now confine ourselves to the canonical observables $\hat{x}$ and $\hat{p}$, the passage from the DCR~(\ref{comm}) to GUP~(\ref{gup}) is as follows:
we first set $\mathcal{\hat{A}} = \hat{x} - \langle \hat{x} \rangle_{\varrho} $ and $\mathcal{\hat{B}} = \hat{p} - \langle \hat{p} \rangle_{\varrho}$ so that $(\Delta{{x}})^2_{\varrho} = \langle \mathcal{\hat{A}}^2\rangle_{\varrho} $, $(\Delta{{p}})^2_{\varrho} = \langle\mathcal{\hat{B}}^2\rangle_{\varrho} $ and $[\hat{x}, \hat{p}]\hat{\varrho} = [\mathcal{\hat{A}},\mathcal{\hat{B}}]\hat{\varrho}$; then, for an arbitrary vector $\psi \in {\mbox{Ran}} \hat{\varrho}$ and a generic $\gamma \in \mathbb{R}$, we have
\begin{eqnarray}
0 &\leq& |\!|(\mathcal{\hat{B}} -i\gamma \mathcal{\hat{A}})\psi  |\!|^2 \nonumber \\[2mm]
&=& \langle \psi|  \mathcal{\hat{B}}^2 | \psi \rangle + i\gamma \langle \psi | [\mathcal{\hat{A}},\mathcal{\hat{B}}]| \psi \rangle + \gamma^2 \langle \psi|  \mathcal{\hat{A}}^2 | \psi \rangle\, .
\label{4.ab}
\end{eqnarray}
Thus,
\begin{eqnarray}
\mbox{Tr}(\mathcal{\hat{B}}^2 \hat{\varrho}) + i\gamma \mbox{Tr}([\mathcal{\hat{A}},\mathcal{\hat{B}}]\hat{\varrho})  + \gamma^2 \mbox{Tr}(\mathcal{\hat{A}}^2 \hat{\varrho}) \ \geq \ 0\, .
\label{5.aac}
\end{eqnarray}
The LHS is smallest for $\gamma =  i \mbox{Tr}([\mathcal{\hat{B}},\mathcal{\hat{A}}]\hat{\varrho}) / (2 \mbox{Tr}(\mathcal{\hat{A}}^2 \hat{\varrho}))$, which brings (\ref{5.aac}) to the form
\begin{eqnarray}
\mbox{\hspace{-3mm}}\mbox{Tr}(\mathcal{\hat{A}}^2 \hat{\varrho}) \mbox{Tr}(\mathcal{\hat{B}}^2 \hat{\varrho})   =  (\Delta{{x}})^2_{\varrho} (\Delta{{p}})^2_{\varrho} \ \geq \ \frac{1}{4} \mbox{Tr}( i[\hat{x},\hat{p}]\hat{\varrho})^2.
\end{eqnarray}
This is nothing but the quantum mechanical version of the {\em covariance inequality} known from probability calculus
~\cite{Muk}.
By employing~(\ref{comm}), we now obtain
\begin{eqnarray}
(\Delta{{x}})_{\varrho} (\Delta{{p}})_{\varrho} \ \geq \  \frac{\hbar}{2}\lf(1+\beta\,\frac{(\Delta{{p}})^2_{\varrho} + \langle \hat{p} \rangle^2_{\varrho}  }{m_p^2}\ri).
\label{split}
\end{eqnarray}
If the density matrix $\varrho$ is {\em mirror symmetric} (i.e.,  $\langle \hat{p} \rangle_{\varrho} = 0$), then 
the inequality~(\ref{split}) indeed coincides with the GUP~(\ref{gup}).


To find  $\hat{\varrho}$ that saturates the GUP (\ref{split}), we observe from~(\ref{4.ab}) that the inequality is saturated if and only if for all $\psi \in {\mbox{Ran}}\hat{\varrho}$  the equation $(\mathcal{\hat{B}} -i\gamma \mathcal{\hat{A}})|\psi \rangle = 0$ holds. If this equation has more than one solution for given $\gamma$, $\langle \hat{x} \rangle_{\varrho}$ and $\langle \hat{p} \rangle_{\varrho}$,
the corresponding $\varrho$ with the minimum uncertainty is a mixture of CSs (i.e., pure GUP-saturating states).
%
%
Unless otherwise specified, we will further assume that $\langle \hat{x} \rangle_{\varrho} = x_0 = 0$, since $x_0$ will only affect the phase factor in CSs (see Section~\ref{II.G.cc}).
%

It is apparent that on the class of mirror symmetric $\varrho$'s, the equation
%
%
%
%
\be\label{eqn}
\lf(\hat{p}-i\gamma \hat{x}\ri)|\psi\rangle \ = \ 0\,,
\ee
admits only one solution for $\psi \in L^2(\mathbb{R})$ [cf., e.g., Eq.~(\ref{diffeq})], so that the minimum–uncertainty $\hat{\varrho}$  is a pure (coherent) state.
It is convenient to seek the solution to (\ref{eqn}) in the momentum representation, \emph{i.e}, $|\psi\rangle \mapsto  \psi(p)=\langle p|\psi\rangle$.
In the momentum space, $\hat{x}$ and $\hat{p}$ satisfying DCR can be represented as~\cite{Kempf}
\begin{eqnarray}
&&\hat{p}\,\psi(p)\ = \ p\,\psi(p)\,, \nonumber \\[2mm]
&&\hat{x}\,\psi(p)\ = \ i\hbar\lf(1+\beta\,\frac{p^2}{m_p^2}\ri)\frac{d}{d p}\psi(p)\,.
\label{operatori}
\end{eqnarray}
However, in doing so, the non-symmetric nature of $\hat{x}$  would produce an inconsistent variance for the ensuing CS.
%
For this reason, we resort to another representation of $\hat{x}$ and $\hat{p}$ complying with (\ref{comm}), namely
\begin{eqnarray}
&&\mbox{\hspace{-7mm}}\hat{p}\,\psi(p)\ = \ p\,\psi(p)\,, \nonumber \\[2mm]
&&\mbox{\hspace{-7mm}}\hat{x}\,\psi(p)\ = \ i\hbar\lf(\frac{d}{d p}+\frac{\beta}{2\,m_p^2}\,\ \left\{ p^2,\frac{d}{d p}\right\}\ri)\psi(p)\,,
\label{operators}
\end{eqnarray}
with $\{~,~\}$ representing the anticommutator (more details on the self-adjointness of this operator are discussed in the accompanying letter~\cite{JLLP}).
With this, we  can cast~(\ref{eqn}) into an equivalent form
\be\label{diffeq}
\frac{d}{d p}\,\psi(p) \ = \ -\frac{\lf(1+\frac{\beta\ga\hbar}{m_p^2}\ri)}{\ga\hbar\lf(1  +  \beta\frac{p^2}{m_p^2}\ri)}\,p\,\psi(p)\,,
\ee
whose generic solution is
\begin{eqnarray}
\psi(p) \ = \ N\left[1  +  ({\beta\,p^2})/{m_p^2}\right]_{+}^{-\frac{m_p^2}{2\beta\ga\hbar}-\frac{1}{2}}\, ,
\label{sol}
\end{eqnarray}
which represents a two-parameter class  of CSs with parameters $\beta$ and $\gamma$. Here,  $[z]_{+} = \max\{z,0 \}$, which guarantees that the wave functions (\ref{sol}) are single-valued.
At this stage, it is important to distinguish two qualitatively different situations, namely
$\beta >0$ and $\beta <0$. 

\subsection{Positive $\beta$ case}
\label{betapos}

The coefficient $N$ in (\ref{sol}) ensures that $\int|\psi(p)|^2dp=1$ and for $\beta >0$
%
\be\label{normal}
N_{_>} \ = \ \sqrt{\sqrt{\frac{\beta}{m^2_p\pi}}\,\frac{\Gamma\lf(\frac{m_p^2}{\beta\ga\hbar}+ 1\ri)}{\Gamma\lf(\frac{m_p^2}{\beta\ga\hbar} + \frac{1}{2}\ri)}}\,,
\ee
with $\Gamma(x)$ being the conventional Euler's gamma function.
\begin{center}
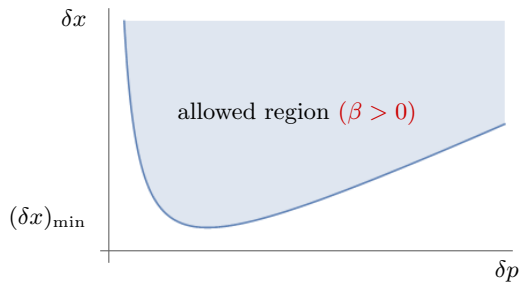
\begin{figure}[h]
\hspace{12mm}\includegraphics[width=5.5cm]{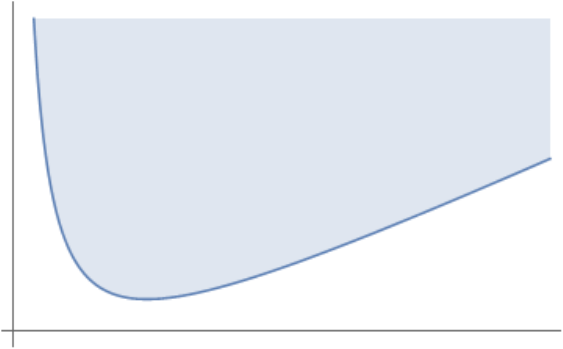}
\begin{picture}(10,4)
\put(-174,90){{$\delta x$} } \put(-10,-5){{$\delta p$} } \put(-130,55){{allowed region {\cdred($\beta > 0$)}}} \put(-194,15){$(\delta x)_{\rm{min}}$}
\end{picture}\\[4mm]
\caption{\footnotesize{For $\beta > 0$ the GUP implies a ``minimal length'' $(\delta x)_{\rm min} = \hbar \sqrt{\beta }/m_p$.}}\label{Fig1}
\end{figure}
\end{center}
It should be stressed that the GUP inequality (\ref{gup})  with $\beta>0$ predicts the existence of a unique minimal $\delta x$, such that 
\begin{eqnarray}
(\delta x)_{\rm min} \ = \ \hbar \sqrt{\beta }/m_p \ = \ \sqrt{\beta } \ \! \ell_p\, ,
\end{eqnarray}
where $\ell_p = \sqrt{\hbar G/c^3}\approx 10^{-33}$cm is the Planck length (see Fig.~\ref{Fig1}). 
As argued in Ref.~\cite{Kempf}, a minimal uncertainty in position means that the position operator is no longer
essentially self-adjoint, but only symmetric. As a matter of fact, one might appropriately restrict the definition region of $\hat{x}$ so that the deficiency index 
is $(0,0)$, which consequently leads to a one-parameter class of self-adjoint extensions of the operator $\hat{x}$ on $L^2(\mathbb{R})$, cf. also Ref.~\cite{JLLP}. Unfortunately, the spectrum of such $\hat{x}$ operators is discrete, which is not compatible with the $\beta >0$ case. Indeed, the expectation value of the deformed commutation relation~\eqref{comm} with respect to any eigenstate of $\hat{x}$ gives a zero left-hand side of~\eqref{comm}, while the right-hand side is always non-zero for $\beta>0$. Such a situation would not happen should both $\hat{p}$ and $\hat{x}$ have a continuous spectrum, because the corresponding eigenstates do not belong to the
domain of the commutator.
On the other hand, the expectation value of the right-hand side of Eq.~\eqref{comm} can be zero but only if $\beta < 0$, which however would contradict our original assumption $\beta >0$. 
A rescue strategy that is often employed
in this context~\cite{Kempf} is to keep $\hat{x}$ being merely symmetric (which ensures that all expectation
values are real) while giving up on self-adjointness. In turn, this opens the way
for the introduction of minimal positional uncertainties.

The aforementioned mismatch between 
a discrete spectrum of the operator $\hat{x}$ and the existence of $(\delta x)_{\rm min}$ leads to yet another interesting point.  
In contrast to conventional 
lattice discretization of space, the aforementioned $(\delta x)_{\rm min}$ does not introduce any UV cutoff scale in the momentum space that would be proportional to the inverse lattice spacing. In other words, the GUP with $\beta>0$ does not provide any universal upper bound for $\delta p$, which instead naturally appears in the GUP with negative $\beta$ (see Fig.~\ref{Fig2}).

\subsection{Negative $\beta$ case}
\label{betaneg}

In this case, the CS (\ref{sol}) can be rewritten in the following, equivalent form
\begin{eqnarray}
\psi(p) \ = \ N_{_<}\lf[1  - ({|\beta|\,p^2})/{m_p^2}\ri]_{+}^{\frac{m_p^2}{2|\beta|\ga\hbar}-\frac{1}{2}}\, ,
\label{sol2}
\end{eqnarray}
where 
\begin{eqnarray}
N_{_<} \ = \ \sqrt{\sqrt{\frac{|\beta|}{m^2_p\pi}}\,\frac{\Gamma\lf(\frac{1}{2} + \frac{m_p^2}{|\beta|\ga\hbar}\ri)}{\Gamma\lf(\frac{m_p^2}{|\beta|\ga\hbar}\ri)}}.
\end{eqnarray}
The physical scenario with $\beta < 0$ has been less explored in literature than the $\beta > 0$ case,
though the related GUP has a number of relevant implications in
cosmology~\cite{JKS,CGgrav}, astrophysics~\cite{Ong} and DSR~\cite{MS1,MS2} and it is also mathematically better behaved.
Note that for $\beta < 0$,  Eq.~(\ref{sol}) involves noninteger power of negative reals, which generally leads to multi-valued CS. Because wave functions must be single-valued, CS has to have bounded support, which means that $\hat{p}$ must be bounded with spectrum $|\sigma(\hat{p})|\leq m_p/\sqrt{|\beta|}$. The ensuing operator $\hat{x}$ corresponding to the formal differential expression (\ref{operators}) is self-adjoint and has a continuous spectrum~\cite{JLLP}.  Consequently, none of the eigenvectors of $\hat{x}$ belong to the Hilbert space $L^2((-m_p/\sqrt{|\beta|},m_p/\sqrt{|\beta|}))$ (just like none of the eigenvectors of $-i\hbar \nabla$ belong to the Hilbert space $L^2(\mathbb{R}^d)$). 
Instead, they belong to the space ${{\mathcal{S}}}'((-m_p/\sqrt{|\beta|},m_p/\sqrt{|\beta|}))$, namely the space of complex-valued tempered distributions.
Hence, there is a spectral transition from discrete to continuous spectrum when $\beta$ becomes negative.
\begin{center}
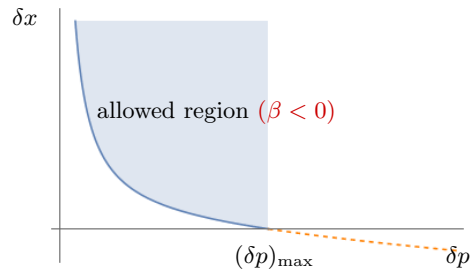
\begin{figure}[h]
\hspace{12mm}\includegraphics[width=5.5cm]{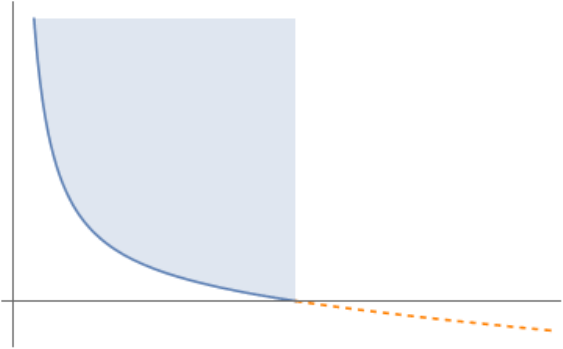}
\begin{picture}(10,4)
\put(-174,90){{$\delta x$} } \put(-10,-0){{$\delta p$} } \put(-142,55){{allowed region {\cdred($\beta < 0$)}}} \put(-90,-0){$(\delta p)_{\rm{max}}$}
\end{picture}
\\[4mm]
\caption{\footnotesize{For $\beta < 0$ the GUP implies  a ``maximal momenta'' $(\delta p)_{\rm max} = m_p/\sqrt{|\beta|}$.}}\label{Fig2}
\end{figure}
\end{center}
In passing, we pinpoint that, since eigenstates are now outside of the Hilbert space, one can avoid subtleties with 
deformed commutation relations encountered in the $\beta>0$
case. 

As a final remark, let us observe that as $\beta \rightarrow 0$, both~(\ref{sol}) and (\ref{sol2}) reduce to the usual minimum-uncertainty Gaussian wave-packet (Glauber coherent state)
associated with the conventional Heisenberg uncertainty relations.

\subsection{Physical meaning of $\gamma$ parameter}
\label{gamma}

While the role of $\beta$ as the GUP deformation parameter is quite clear, the role of $\gamma$ is less obvious. 
In order to find a meaning for $\gamma$, we note that (\ref{eqn})
implies
\begin{eqnarray}
0 &=& \langle \psi |\lf(\hat{p}+i\gamma \hat{x}\ri)\lf(\hat{p}-i\gamma \hat{x}\ri)|\psi\rangle \nonumber \\[2mm]
&=& (\Delta p)^2  +  i\gamma \langle \psi | \lf[\hat{x},\hat{p}\ri]|\psi\rangle  +
 \gamma^2 (\Delta x)^2
 \nonumber \\[2mm]
&=& (\Delta p)^2  -  \gamma \ \! |\langle \psi | \lf[\hat{x},\hat{p}\ri]|\psi\rangle|  +  \gamma^2 \frac{|\langle \psi | \lf[\hat{x},\hat{p}\ri]|\psi\rangle|^2 }{4(\Delta p)^2}\, ,~~~~
 \end{eqnarray}
where in the last line we used the fact that (\ref{split}) is saturated. The above expression has a single solution for $\gamma$, which
with the help of (\ref{comm}) can be cast as
\begin{eqnarray}
\gamma \ = \ \frac{2(\Delta p)^2}{\hbar\lf[1 \ + \ \beta\,{(\Delta p)^2}/{m_p^2}\ri]}\,.
\label{13cd}
\end{eqnarray}
By realizing that (\ref{13cd}) is valid for CSs, we can rewrite it in a more succinct form as
\begin{eqnarray}
\gamma^2 \ = \ \frac{(\Delta p)^2_{CS}}{(\Delta x)^2_{CS}}\, .
\label{18cd}
\end{eqnarray}
We observe that  $\gamma$ can be defined also in the limit $\Delta p \rightarrow \infty$, even though GUP (\ref{split}) is in such a case meaningless. 
Note also that CSs (\ref{sol})  are indeed mirror symmetric states and they satisfy   $\langle \hat{p} \rangle_{CS} = \langle \hat{x} \rangle_{CS} = 0$. The relation (\ref{13cd}) will play an important role in Section~\ref{TSDist}.

\subsection{Another look at the CS (\ref{sol})}
\label{tsallis 1a}

To better understand the structure of (\ref{sol}), we rephrase the two-parameter class of CSs 
in terms of another two
parameters (say $q$ and $b$) so that 
\begin{eqnarray}\label{qb}
q \ = \ \frac{\beta \gamma \hbar}{m^2_p+\beta\ga\hbar} + 1\,, \qquad  b \ = \ \frac{2 m_p}{\gamma \hbar}+\frac{2\beta}{m_p}\, .
\label{17.ccd}
\end{eqnarray}
This relation is supposed to be valid for $\beta \lessgtr 0$. In particular, we can observe with the help of (\ref{13cd}) that, for fixed variance  $\Delta p$, the $q$ parameter is a monotonically increasing  function of $\beta$ provided $\beta \neq -m_p^2/[3 (\Delta p)^2]$, in which case the value of $q$ is undefined.
Should we have no deformation (i.e., $\beta = 0$), then
$q=1$.

The previous reparametrization  allows us to set (\ref{sol}) in the form 
\begin{eqnarray}\label{finalsol}
\psi(p) \ = \ {N}_{_{\lessgtr}} \left[1 - b\,(1-q) \frac{p^2}{2 m_p} \right]^{\frac{1}{2(1-q)}}_+\, .
\label{CS-1}
\end{eqnarray}
This is nothing but the probability amplitude for the
Tsallis distribution of a free non-relativistic particle
\begin{eqnarray}
\mbox{\hspace{-2mm}}q_{_{T}}(p|q,b)  =   |\psi(p)|^2  =  \frac{1}{Z}\left[1  -  b\,(1-q) \frac{p^2}{2 m_p}\right]^{\frac{1}{1-q}}_+\!\! ,
\label{15aa}
\end{eqnarray}
with $Z= {N}_{_{\lessgtr}}^{-2}$ being the ``partition function''.

\subsection{Toward a relativistic generalization}
\label{relat}

It is interesting to point out that Eq.~(\ref{sol}) can be brought also into a relativistic-like form. To see this,
we perform another substitution, namely
\begin{eqnarray}
q \ = \ \frac{\beta \gamma \hbar}{m^2_p+\beta\ga\hbar} + 1\,, \;\;\;
\tilde{b} \ = \ \frac{b\,m_p}{1 + b\,(1-q)\,m_p}\,.
\label{24.cc}
\end{eqnarray}
This choice yields
\begin{eqnarray}
\mbox{\hspace{-2mm}}\psi(p) 
&=& {\tilde{N}}_1 \left[1 - \tilde{b}(1-q) \left(\frac{p^2}{2 m^2_p} + 1\right) \right]^{\frac{1}{2(1-q)}} \,.
\label{17bb}
\end{eqnarray}
The last relation is particularly useful, since we can identify it with the leading-order approximation of the function
\begin{eqnarray}
\mbox{\hspace{-5mm}}\psi(p) &\sim& \tilde{N}_2 \left[1 - \tilde{b}\,(1-q) \sqrt{1+ \frac{p^2}{m^2_p}} \right]^{\frac{1}{2(1-q)}}\nonumber \\[3mm]
&\stackrel{q\rightarrow 1}{\longrightarrow} & \tilde{N}_3 \exp\left(- \frac{1}{2 \theta}\,\sqrt{1+ \frac{p^2}{m^2_p}}\right),
\end{eqnarray}
for $p^2 \ll m_p^2$. The corresponding probability distribution
\begin{eqnarray}
\mbox{\hspace{-5mm}}q_J(p|\theta) &=&  |\psi(p)|^2\nonumber \\
&\approx& \frac{1}{2\,m_p K_{1}(1/\theta)}\ \! \exp\left(- \frac{1}{\theta}\,\sqrt{1+ \frac{p^2}{m^2_p}}\right)\!.
\end{eqnarray}
is nothing but the Maxwell--J\"{u}ttner distribution for relativistic particles, with
$\theta \ = \ {kT}/{m_p} \ = \ \gamma\hbar/{(2m_p^2)}$
being the temperature in Planck units. Here and in subsequent reasonings we adopt the convention that $c=1$. At this point, we note that:
\begin{itemize}
  \item should we have considered 3D momentum, then the normalization factor would involve the Bessel function $K_2$ instead of $K_1$;
  \item should we have required the Maxwell--J\"{u}ttner probability amplitude to be an exact coherent state of the GUP, we would need to modify Eq.~(\ref{gup}). The present form of the GUP is only the leading-order approximation towards Maxwell--J\"{u}ttner type of coherent states.
\end{itemize}
In the following, we will focus only on the form of GUP CSs~(\ref{15aa}). GUP with exact Maxwell--J\"{u}ttner type of CSs will be discussed in our future work (see also the discussion in Sec.~\ref{Conc} for more comments on this issue).

\subsection{Some essentials about Tsallis distribution}
\label{TSDist}

A few remarks concerning (\ref{15aa}) are now in order. Tsallis distribution of this type is also known as $q$-Gaussian distribution and denoted as $\exp_q(-b\,p^2/2\,m_p)$.
In the limit $q\rightarrow 1$, $\exp_q(-b\,p^2/2\,m_p) \rightarrow \exp(-b\,p^2/2\,m_p)$.  We have already pointed this out, since $q\rightarrow 1$ is equivalent to $\beta \rightarrow 0$. In addition, because the momentum is unbounded for $q\ge 1$ (i.e. $\beta \ge 0$), the distribution~(\ref{15aa}) is normalizable only for values of $1\leq  q<3$. For values $q<1$ (which means $\beta < 0$), the $q$-Gaussian distribution needs to be set to zero for $p>\sqrt{2\,m_p/b\,(1-q)}$, as only in such cases the argument in the square brackets of Eq.~(\ref{15aa}) is not negative.
Therefore,  for $\beta <0$  the corresponding coherent states must have a finite support. 

Moreover,  for $q \geq 5/3$ the variance of (\ref{15aa}) is undefined (infinite)~ and thus the GUP cannot even be formulated. However, when $q < 5/3$, then
$(\Delta p)^2=2\,m_p/b\,(5-3q)$ (see for example Ref.~\cite{Thistleton}). The latter identity implies that  the parameter $\gamma$ and $(\Delta p)^2$ must be related, namely
\begin{eqnarray}
\mbox{\hspace{-3mm}}(\Delta p)^2  =  \frac{2\,m_p}{b\,(5-3q)}  \  \Rightarrow   \  \gamma  =  \frac{2(\Delta p)^2}{\hbar\lf[1 \ + \ \beta{(\Delta p)^2}/{m_p^2}\ \!\ri]}\,,
\label{var2}
\end{eqnarray}
which precisely coincides with the result (\ref{13cd}). This, in turn, justifies our choice of the representation of $\hat{x}$ and $\hat{p}$ operators.
Should we have started with the representation (\ref{operatori}) instead, then we would have arrived at the
Tsallis-type coherent states with slightly different parameters than in (\ref{sol}). The ensuing variance would have then implied
\begin{eqnarray}
\gamma  \ = \  \frac{2(\Delta p)^2}{\hbar\lf[1 \ + \ 3\beta{(\Delta p)^2}/{m_p^2}\ \!\ri]}\,,
\label{19.bbb}
\end{eqnarray}
which is clearly incompatible with (\ref{13cd}).

Furthermore, it should be stressed that the mean value does not exist for $q>2$, so such coherent states cannot be mirror symmetric as assumed.
In the upcoming considerations, we will thus restrict our attention to the physical domain of interest $q < 5/3$ with a particular focus on the negative $\beta$, which entails $q<1$.

In passing, we emphasize that the Tsallis distribution is equivalent to the Student-$t$ distribution, which is used in statistical hypothesis testing, but it also arises
naturally in situations with small sample sizes (namely $2/(1-q)$) or samples with unknown standard deviations.

\subsection{The connection of states (\ref{CS-1}) with generalized coherent states}\label{II.G.cc}

Canonical (or Glauber's) CSs known from the (multimode) quantum harmonic oscillator satisfy three basic properties~\cite{Mandel,BJV,perelomov}: 1) they are eigenstate of lowering operators, 2) they are generated via translation (or displacement) operators and 3) they are minimum uncertainty states.  There exists a huge variety of the so called generalized CSs, which maintain only some of the above three conditions.  For instance, generalizations based on translation operators were developed mostly by Perelomov~\cite{perelomov} for systems with group-related fundamental commutation relations (hence the name group-related or Perelomov CSs). Coherent states which are eigenstates of lowering operators were generalized mostly by Barut and Girardello~\cite{BG}, again in the group-theoretic context (hence the name Barut--Girardello's CSs).  Finally, CSs that saturate uncertainty relations were generalized e.g. by Nieto {\em et al.}~\cite{nieto} by following an analogy with Schrödinger’s original definition of CSs (hence they are called Nieto's or Schrödinger-Nieto's CSs). 
Accordingly, our CSs belong conceptually to the same class of generalized CSs as the Schrödinger--Nieto CSs.

From a mathematical-physics point of view, any set of would-be CSs qualifies  as a {\em family of (generalized) coherent states}  if it meets the following two properties~\cite{KlauderIIa}:
\begin{itemize}
  \item all elements of the set are strongly continuous functions of their label variables;
\item there exists a measure on the label space such that the unit operator admits the resolution of unity.
\end{itemize}
We will now demonstrate that the Tsallis probability amplitudes (\ref{sol}) can be incorporated into a set of generalised CSs that saturate the uncertainty relation (\ref{split}).
%
%
%

Without assuming mirror symmetry and the auxiliary constraint $\langle \hat{x} \rangle_{\psi} = x_0 = 0$, we observe that (\ref{split}) is saturated by states satisfying 
\begin{eqnarray}
(\mathcal{\hat{B}} -i\gamma \mathcal{\hat{A}})|\psi \rangle \ = \ \left[\hat{p} -p_0-i\gamma (\hat{x} -x_0)\right]|\psi\rangle \ = \ 0\, ,
\label{30.cd}
\end{eqnarray}
where $p_0$ and $x_0$ are  expectation values of $\hat{p}$ and $\hat{x}$ in the state $|\psi\rangle$, respectively. The corresponding differential equation has the solution (for concreteness we focus on the $\beta<0$ case)
\begin{eqnarray}
&&\mbox{\hspace{-10mm}}\psi(p; p_0,x_0) \ \equiv \ \langle p|\psi, p_0,x_0\rangle  \nonumber \\[2mm]
&&\mbox{\hspace{-10mm}}=  N e^{-\frac{\varpi m_p}{\hbar \ga \sqrt{|\beta|}} \mbox{\footnotesize{arctanh}}(p\sqrt{|\beta|}/m_p) }\left[1  -  \frac{{|\beta|\,p^2}}{m_p^2}\right]_{+}^{\frac{m_p^2}{2|\beta|\ga\hbar}-\frac{1}{2}}\!,~
\label{31.cdf}
\end{eqnarray}
where $\varpi = i\gamma x_0 - p_0$ and $N$ is a normalization constant. 

Two comments are now in order: a) the actual value of $x_0$ in (\ref{31.cdf}) does not affect the probability distribution (\ref{15aa}), as it only appears in the phase factor, and b) the parameter $\gamma$ in (\ref{30.cd})-(\ref{31.cdf})   is now explicitly dependent on $p_0$. As for the last point, by following the same procedure as in Section~\ref{gamma}, we arrive at 
\begin{eqnarray}
\gamma  =  \frac{2(\Delta p)^2}{\hbar\lf\{1 \ + \ \beta{\left[(\Delta p)^2 + p_0^2\right]}/{m_p^2}\ \!\ri\}} \ = \ \frac{\Delta p}{\Delta x}\,.
\label{32.cde}
\end{eqnarray}
The latter can be written formally as $(\Delta p)_{CS}/(\Delta x)_{CS}$, which emphasises that the variances have been calculated with respect to states that satisfy the GUP (\ref{split}).

When we adhere to the usual HUP and fix  $(\Delta p)_{CS}$ and $(\Delta x)_{CS}$, we can see that $\gamma$ is fixed  while the expectation value of momenta, $\langle \hat{p}\rangle_{\psi_{CS}} = p_0$, is unrestricted  (it does not appear explicitly in HUP).
In the present case when we fix $(\Delta p)_{CS}$ and $(\Delta x)_{CS}$, we still get fixed  $\gamma$ but due to (\ref{split}) $p_0$ is not free anymore --- it is fixed as well. This makes it difficult to define measure on the label space $\{x_0,p_0\}$ that would in the limit $|\beta| \rightarrow 0$ converge to Glauber's measure. So, rather than fixing  $(\Delta p)_{CS}$ and $(\Delta x)_{CS}$, we fix instead directly $\gamma$  (which fixes only fraction of $(\Delta p)_{CS}$ and $(\Delta x)_{CS}$ but not their respective values) in which case the GUP (\ref{split}) does not restrict $p_0$ to some specific value. In fact, it will be shortly seen that this strategy will allow us to define a measure on the label space $\{x_0,p_0\}$ with a correct limiting behavior.

We note, in passing, that in the limit $|\beta| \rightarrow 0$, Eq.~(\ref{31.cdf}) reduces to the usual minimal uncertainty Schr\"{o}dinger wave packet
\begin{eqnarray}
&&\mbox{\hspace{-5mm}}\psi(p; p_0,x_0)|_{\beta \rightarrow 0} \ \propto \ \exp\left[\frac{i x_0(p_0-p)}{\hbar} - \frac{(p - p_0)^2}{2\gamma \hbar}\right]\nonumber \\[2mm]
&&\mbox{\hspace{5mm}}= \
\exp\left[\frac{i x_0(p_0-p)}{\hbar} - \frac{(p - p_0)^2}{4(\Delta p)^2}\right] ,~~~~~~
\end{eqnarray}
where in the second line we used the fact that $\gamma|_{|\beta| \rightarrow 0} = 2(\Delta p)^2/\hbar \equiv 2(\Delta p)^2_{CS}/\hbar $ [cf.,  Eq.~(\ref{32.cde})].

When $\gamma$ is fixed, it is not difficult to see that states (\ref{31.cdf}) are strongly continuous functions of the label variables $x_0$ and $p_0$.  A strong continuity means~\cite{KlauderIIa} that for every convergent label set such that $\{p_0',x_0'\} \rightarrow \{p_0,x_0\}$, the distance between the two corresponding quantum states  $|\!| |\psi, p_0',x_0'\rangle -  |\psi, p_0,x_0\rangle  |\!| \rightarrow 0$, with $|\!| |\psi \rangle |\!|   = \sqrt{\langle \psi|\psi \rangle }$. 

Now, since 
\begin{eqnarray}
&&\mbox{\hspace{-15mm}}|\!| |\psi, p_0',x_0'\rangle -  |\psi, p_0,x_0\rangle  |\!|^2 \nonumber \\[2mm] 
&&= \ |\!| |\psi, p_0',x_0'\rangle |\!|^2 + |\!| |\psi, p_0,x_0\rangle |\!|^2 \nonumber \\[2mm]
&&\mbox{\hspace{3mm}}- 2\mbox{Re} \langle \psi, p_0',x_0'|\psi, p_0,x_0\rangle\, ,
\end{eqnarray}
the strong continuity of the vectors is thus a simple consequence of the  continuity 
of the matrix element of any two CSs in their label variables. In the present case
we have
\begin{eqnarray}
&&\mbox{\hspace{-4mm}}\langle \psi, p_0',x_0'|\psi, p_0,x_0\rangle \nonumber\\[2mm] 
&&\mbox{\hspace{-2mm}}= \ |N|^2\int_{-m/\sqrt{|\beta|}}^{m/\sqrt{|\beta|}} dp \ \! e^{-\frac{(\varpi'^* + \varpi) m_p}{\hbar \ga \sqrt{|\beta|}} \ \! \mbox{\footnotesize{arctanh}}(p\sqrt{|\beta|}/m_p) }\nonumber \\[2mm]
&&\mbox{\hspace{-1mm}}\times \left[1  -  \frac{{|\beta|\,p^2}}{m_p^2}\right]_{+}^{\frac{m_p^2}{|\beta|\ga\hbar}-1}\nonumber \\[2mm]
&&\mbox{\hspace{-2mm}}= \ |N|^2 \frac{m_p}{\sqrt{|\beta|}}\int^{\infty}_{-\infty} dz \ \!
 e^{-\frac{(\varpi'^* + \varpi) m_p}{\hbar \ga \sqrt{|\beta|}} z} \ \! \left(\cosh z\right)^{-a}\! .~~~~~
\end{eqnarray}
Here $|N|^2$ is a shorthand notation for $N(p_0)N^*(p_0')$ and $a = \frac{2m_p^2}{\hbar \gamma|\beta|}$. Since $\left(\cosh z\right)^{-a}$ is absolutely  integrable over ${\mathbb{R}}$ and decreases at infinity faster than any power of $|z|^{-1}$ [in fact faster than 
$\exp(-|z| a)$], we may differentiate under the integral sign any number of times. By virtue of the dominated convergence theorem, we can write for  $|p_0|, |p_0'| < m_p/\sqrt{|\beta|}$

\begin{widetext}
\begin{eqnarray}
\mbox{\hspace{-4mm}}\langle \psi, p_0',x_0'|\psi, p_0,x_0\rangle &=& |N|^2 \frac{m_p}{\sqrt{|\beta|}} e^{i\frac{(p_0'+p_0)m_p}{\hbar \gamma \sqrt{|\beta|}} \frac{\partial }{ \partial y}}\left. \int_{-\infty}^{\infty} dz \ \! e^{iy z}   \ \! \left(\cosh z\right)^{-a}\right|_{y = \frac{(x_0'-x_0)m_p}{\hbar \sqrt{|\beta|}}} \nonumber \\[2mm]
&=& |N|^2 \frac{m_p}{\sqrt{|\beta|}} e^{i\frac{(p_0'+p_0)m_p}{\hbar \gamma \sqrt{|\beta|}} \frac{\partial }{ \partial y}} \left.\left( \frac{2^{a -1}}{\Gamma(a)}  \left|\Gamma\left(\frac{a}{2} + i \frac{y}{2}\right)\right|^2 \right)\right|_{y = \frac{(x_0'-x_0)m_p}{\hbar \sqrt{|\beta|}}}\nonumber \\[2mm]
&=& \frac{\Gamma\left(\frac{a}{2} - \frac{(p_0'+p_0)m_p}{2 \hbar \gamma \sqrt{|\beta|}} + i \frac{(x_0'-x_0)m_p }{2 \hbar \sqrt{|\beta|}}\right)\Gamma\left(\frac{a}{2} + \frac{(p_0'+p_0)m_p}{2 \hbar \gamma \sqrt{|\beta|}} - i \frac{(x_0'-x_0)m_p }{2 \hbar \sqrt{|\beta|}}\right)}{\sqrt{\Gamma\left(\frac{a}{2} - y \right)\Gamma\left(\frac{a}{2} + y\right)} \sqrt{\Gamma\left(\frac{a}{2} - y' \right)\Gamma\left(\frac{a}{2} + y' \right)}}\, .~~~
\end{eqnarray}
\end{widetext}
Here $y = \frac{p_0m_p}{\hbar \gamma \sqrt{|\beta|}}$ and $y' = \frac{p_0'm_p}{\hbar \gamma \sqrt{|\beta|}}$.
In the second line we have used  the Ramanujan formula~\cite{Ramanujan}. 

We now employ the fact that
for $\mbox{Re}(\zeta) > 0$, the gamma function $\Gamma(\zeta)$ is a continuous function of its argument and satisfies $[\Gamma(\zeta)]^{*} = \Gamma(\zeta^{*})$. This, in turn, implies that the matrix elements of any two CSs are continuous in their label variables.
The latter is particularly true  for a subclass of states with $p_0 = 0$, 
i.e., mirror symmetric states.


To prove the existence of the integral measure $\mu(x_0,p_0)$ for the resolution of unity, we need to show that there exists $\mu(x_0,p_0)$ such that
\begin{widetext}
\begin{eqnarray}
\int_{-\infty}^{\infty} dx_0 \int_{-m_p/\sqrt{|\beta|}}^{m_p/\sqrt{|\beta|}} dp_0 \ \! \mu(x_0,p_0) \ \! \langle p'|\psi, p_0,x_0\rangle\langle \psi, x_0, p_0|p\rangle \ = \ \delta(p' - p)\, .
\label{35.cfg}
\end{eqnarray}
%
%
The LHS of (\ref{35.cfg}) can be explicitly written in the form
\begin{eqnarray}
&&\int_{-\infty}^{\infty} dx_0 \int_{-m_p/\sqrt{|\beta|}}^{m_p/\sqrt{|\beta|}} dp_0 \ \! \mu(x_0,p_0) \ \! |N(p_0)|^2 \ \!e^{i\frac{x_0 m_p}{\hbar  \sqrt{|\beta|}} \ \! \left[ \mbox{\footnotesize{arctanh}}(p'\sqrt{|\beta|}/m_p) - \mbox{\footnotesize{arctanh}}(p\sqrt{|\beta|}/m_p)\right]}\nonumber \\[2mm]
&&\mbox{\hspace{1cm}} \times \ e^{\frac{p_0 m_p}{\hbar \gamma \sqrt{|\beta|}} \ \! \left[ \mbox{\footnotesize{arctanh}}(p'\sqrt{|\beta|}/m_p) + \mbox{\footnotesize{arctanh}}(p\sqrt{|\beta|}/m_p)\right]} \ \! \left[\left(1  -  \frac{{|\beta|\,p'^2}}{m_p^2}\right)\left(1  -  \frac{{|\beta|\,p^2}}{m_p^2}\right)\right]^{\frac{m_p^2}{2|\beta|\ga\hbar}-\frac{1}{2}}\, ,
\label{36.b}
\end{eqnarray}
\end{widetext}
where
\begin{eqnarray}
|N(p_0)|^2 \ = \ \frac{\sqrt{|\beta|}}{m_p} \frac{\Gamma\left( a\right)}{2^{a -1} } \ \! \left[\Gamma\left(\frac{a}{2} - y \right)\Gamma\left(\frac{a}{2} +y \right)\right]^{-1}\! .~~
\label{37.ccf}
\end{eqnarray}
The latter goes to zero when $p_0$ reaches its bounding values $\pm m_p/\sqrt{|\beta|}$.
In passing we note that Stirling's approximation implies the correct limiting behavior
\begin{eqnarray}
|N(p_0)|^2|_{|\beta| \rightarrow 0}   =  \frac{\exp\left(-\frac{p_0^2}{\gamma \hbar}   \right)}{\sqrt{\pi \gamma \hbar }}  =  \frac{\exp\left(-\frac{p_0^2}{2 (\Delta p)^2}   \right)}{\sqrt{2\pi (\Delta p)^2}} \, .~~~
\end{eqnarray}
Since (\ref{36.b}) equals to $\delta(p'-p)$, $\mu(x_0,p_0)$ must depend only on $p_0$. This allows to perform the integration 
over $x_0$ and rewrite (\ref{36.b}) in the form
\begin{eqnarray}
&&\mbox{\hspace{-10mm}}2\pi \hbar \int_{-m_p/\sqrt{|\beta|}}^{m_p/\sqrt{|\beta|}} dp_0 \ \! \mu(p_0) \ \! |N(p_0)|^2 \ \!e^{\frac{2 p_0 m_p}{\hbar \gamma \sqrt{|\beta|}} z} \nonumber \\[2mm] 
&&\times \ \left(\cosh z\right)^{-\frac{2m_p^2}{\hbar \gamma|\beta| }}\! \ \delta(p' - p)\nonumber \\[2mm]
&&\mbox{\hspace{-10mm}}= \ \frac{2\pi
\hbar^2\gamma \sqrt{|\beta|}}{m_p}
\int_{-a/2}^{a/2} dy \ \! {\tilde{\mu}(y)} \ \! |\tilde{N}(y)|^2 \ \!e^{2yz}
\nonumber \\[2mm] 
&&\times \ \left(\cosh z\right)^{-a } \ \! \delta(p' - p)
\, .
\label{40.cf}
\end{eqnarray}
Here we have set $z = \mbox{{arctanh}}(p\sqrt{|\beta|}/m_p)$,  $\tilde{\mu}(y) = \mu(p_0)$ and $\tilde{N}(y) = N(p_0)$. 
So, the measure $\mu(p_0)$ can be obtained by taking the inverse of the finite Laplace transform of $(\cosh z)^a$.
After some analysis (cf. Appendix~C for technical details) we obtain
\begin{eqnarray}
&&\mbox{\hspace{-10mm}}\mu(x_0,p_0) \ = \ \mu(p_0)  \ = \ \tilde{\mu}(y) \nonumber \\[2mm]\label{c.11.cfftext}
&&\mbox{\hspace{-5mm}}= \ \frac{1}{2\pi\hbar} \frac{1}{\left[1-\left(\frac{2y}{a}\right)^{\!2}\right]} \ \! \sum_{k=0}^{\infty} \delta(\pm y + a/2 -k)\, ,
 \label{42.cd}
\end{eqnarray}
with $y \in (-a/2,a/2)$. Both sign conventions are admissible. So, the measure is discrete in the $p_0$ variable while continuous in the $x_0$ variable. 
Note that in (\ref{36.b}) the Cauchy principal value integral should be utilised in the $p_0$ integration in  order to 
see that the endpoint singularities  in (\ref{42.cd}) are integrable.
In Appendix~C we show that in the limit $|\beta| \rightarrow 0$ we regain the conventional (Glauber's) CS measure.  


Let us close this subsection by noting that the UR-saturating CSs  belong to the class of so-called {\em pointer states}, i.e. those states that are least affected by the interaction with the environment~\cite{zurek1,zurek2,zurek3}.
In fact, Schr\"{o}dinger's CSs maximize Shannon--Gibbs entropy 
subject to prior data one
possesses about a system, namely the first two moments of the position and momentum variables. Since the latter indirectly reflect system's environment, the probability distribution assigned is the least prone to losing quantum coherence, compared to other pure-state distributions that fulfill the same prior data.
%
%
Such CSs are particularly pertinent in the quasi-classical domain of quantum theory, as they are maximally predictable (or robust) despite of decoherence~\cite{zurek3,zurek4}.

\section{Non-extensive thermodynamics relations}
\label{Tsallis}

It is clear that the probability distributions (\ref{15aa}) and (\ref{17bb}) decay
asymptotically following a power law rather than an exponential law. If we keep variance and mean as the only statistical observables, power-law type distributions are incompatible with the conventional MEP applied to Shannon--Gibbs entropy.
On the other hand, distributions (\ref{15aa}) and (\ref{17bb}) are maximizers for Tsallis
(differential) entropy (TE)~\cite{Tsallis2,JKZ}, i.e., for entropic functional of the form
%
\begin{eqnarray}
{\mathcal{S}}_q^T(\mathcal{F}) &=& \frac{k_T}{(1-q)}\left(  \int_{\mathbb{R}}d{p}\,\mathcal{F}^{q}(p)  - 1 \right) ,
\label{30cf}
\end{eqnarray}
where $\mathcal{F}$ is a probability density function. The constant $k_T$ may in general depend on $q$, and becomes the Boltzmann constant $k_B \approx 1.380649\times 10^{23}$JK$^{-1}$ in the limit $q\rightarrow 1$.
Moreover, it can easily be deduced (by L'H\^{o}pital's rule) that in the limit $q \rightarrow 1$, the TE tends to the Shannon--Gibbs' entropy
\begin{eqnarray}
{\mathcal{S}}^S(\mathcal{F}) &=& -  k_B\int_{\mathbb{R}}d{p}\ \! \mathcal{F}(p)\ln \mathcal{F}(p)\, .
\end{eqnarray}
The coherent state distribution $\exp_{q}\left(-bp^2/m_p \right)$ is the maximizer of the entropy ${\mathcal{S}}_{2-q}^T(\mathcal{F})$  subject to a second moment constraint~\cite{Tsallis1,Tsallis2,JKZ} 
and it
tends to a Gaussian distribution for $q\rightarrow1$.

Given the importance of TE and NTT in following sections, we now derive the prerequisite NTT relations. 
Although these results have already been addressed in the literature (e.g. Ref.~\cite{Tsallis2}), it s instructive to derive them in an alternative and more systematic manner via Caratheodory's theorem.

We first observe that ${\mathcal{S}}_q^T$ satisfies the following non-additivity law for two independent subsystems (say $A$ and $B$)~\cite{Tsallis1,Tsallis2}:
\begin{eqnarray}
\mbox{\hspace{-2mm}}{\mathcal{S}}_q^T(A,B) \!
&=& \!{\mathcal{S}}_q^T(A) \ + \ {\mathcal{S}}_q^T(B) \ + \ \frac{1-q}{k_T}\ \! {\mathcal{S}}_q^T(A){\mathcal{S}}_q^T(B) \nonumber \\[2mm]
&=& \!{k_T}\!\left[\left(\frac{1}{k_T} {\mathcal{S}}_q^T(A)\right)  \oplus_q  \left(\frac{1}{k_T} {\mathcal{S}}_q^T(B)\right)\right], ~~
\label{23cfg}
\end{eqnarray}
where the symbol $\oplus_q$ denotes the $q$-deformed ``sum'', which is defined as in Ref.~\cite{Tsallis2}
\begin{eqnarray}
x \oplus_q  y  \ = \  x  \ + \   y  \ + \   (1-q) x y\, .
\end{eqnarray}
\noindent
In~(\ref{23cfg}) we have employed the short-hand notation ${\mathcal{S}}_q^T(A)  \equiv {\mathcal{S}}_q^T(\mathcal{F}_A)$
and ${\mathcal{S}}_q^T(A,B) \equiv {\mathcal{S}}_q^T(\mathcal{F}_A \cdot \mathcal{F}_B) $, with $\mathcal{F}_X$ being a probability density  associated with the (continuous) random variable describing the system $X$. 

Let us now briefly outline the passage to  thermodynamics based on Tsallis entropy. 
In particular, we will derive the connection between heat one-form and Tsallis entropy (analogue of the Clausius relation for reversible heat exchange) that will be needed in Section~\ref{Verlinde}. 
To this end, we start by considering two systems ($A$ and $B$) in contact (both thermal and mechanical) with each other. Suppose that these have volumes $V(A)$ and $V(B)$ and internal energies $U_q(A)$ and $U_q(B)$, and that each volume and number of particles (as well as the combined
internal energy and total volume) are fixed. 

\begin{widetext}
In thermodynamic equilibrium, the total entropy ${\mathcal{S}}_q^T(A,B)$ must be maximal. By using (\ref{23cfg}) we thus have  
\begin{eqnarray}
\mbox{\hspace{-8mm}}0 \ = \  d{\mathcal{S}}_q^T(A,B) &=& \left[\left(1 + \frac{1-q}{k_T}{\mathcal{S}}_q^T(B) \right)\left(\frac{\partial {\mathcal{S}}_q^T(A)}{\partial U_q(A)}\right)_{V(A)} - \left(1 + \frac{1-q}{k_T}{\mathcal{S}}_q^T(A) \right)\left(\frac{\partial {\mathcal{S}}_q^T(B)}{\partial U_q(B)}\right)_{V(B)}   \right] dU_q(A) \nonumber \\[2mm]
&+&  \left[\left(1 + \frac{1-q}{k_T}{\mathcal{S}}_q^T(B) \right)\left(\frac{\partial {\mathcal{S}}_q^T(A)}{\partial V(A)}\right)_{U_q(A)} - \left(1 + \frac{1-q}{k_T}{\mathcal{S}}_q^T(A) \right)\left(\frac{\partial {\mathcal{S}}_q^T(B)}{\partial V(B)}\right)_{U_q(B)}   \right] dV(A)\, ,
\label{34.jk}
\end{eqnarray}\\
\end{widetext}
where we have employed the fact that the total internal energy and volume are fixed
\begin{eqnarray}
&&U_q(A,B) \ = \ U_q(A) \ + \ U_q(B) \ = \ \mbox{const.}\, , \label{23cf} \\[2mm]\
&&V(A,B) \ = \ V(A) \ + \ V(B) \ = \ \mbox{const.}\, .
\end{eqnarray}
We have also assumed that TE ${\mathcal{S}}_q^T$ is expressed in terms of entropy's natural state variables, i.e. $U_q$ and $V$.

From (\ref{34.jk}), we obtain the two identities, which reflect the fact that when a system is in {\em thermodynamic} equilibrium, then it is simultaneously in {\em thermal} and {\em mechanical} equilibrium.
The first identity one can be written in the form
\begin{eqnarray}
\frac{k_T \beta(A)}{1+ ((1-q)/k_T) {\mathcal{S}}_q^T(A) } &=& \frac{k_T \beta(B)}{1+ ((1-q)/k_T) {\mathcal{S}}_q^T(B) }\nonumber \\[2mm]
&=&k_T \beta^*\, ,
\label{25cc}
\end{eqnarray}
where (by analogy with conventional extensive thermodynamics) we have defined 
\begin{equation}
k_T\beta \ = \ \left(\frac{\partial {\mathcal{S}}_q^T}{\partial U_q}\right)_{V}\, .
\end{equation}
In connection with (\ref{25cc}), it should be emphasized that the {\em physical temperature} is not $(k_T\beta)^{-1}$ (as would be the case at $q \rightarrow 1$),  but rather 
\begin{eqnarray}
\vartheta \ = \ \frac{1}{k_T \beta^*} \ = \ \left( 1 + \frac{1-q}{k_T} {\mathcal{S}}_q^T  \right) \frac{1}{k_T\beta}\, .
\label{29cf}
\end{eqnarray}
Equation~(\ref{25cc}) basically represents the 0th law of thermodynamics, which ensures that one may  assign the same observable empirical temperature $\vartheta$ to all sub-systems in thermal equilibrium. We will shortly see that the empirical temperature  $\vartheta$ can be identified with the ``absolute'' temperature $T_{\rm{phys}}$.

The second identity can be cast as
\begin{eqnarray}
\frac{\left({\partial {\mathcal{S}}_q^T(A)}/{\partial V(A)}\right)_{U_q(A)}  }{1+ ((1-q)/k_T) {\mathcal{S}}_q^T(A) } &=& \frac{\left({\partial {\mathcal{S}}_q^T(B)}/{\partial V(B)}\right)_{U_q(B)}  }{1+ ((1-q)/k_T) {\mathcal{S}}_q^T(B) }\nonumber \\[2mm]
&=&\frac{p_{\rm{phys}}}{\vartheta}\, ,
\label{29ccg}
\end{eqnarray}
Eq.~(\ref{29ccg}) reflects the fact that when two systems are in mechanical equilibrium their pressures are the same. This allows to identify {\em physical pressure} $p_{\rm{phys}}$ as
\begin{eqnarray}
p_{\rm{phys}} \ = \ \frac{\vartheta}{1+ ((1-q)/k_T) {\mathcal{S}}_q^T } \left(\frac{\partial {\mathcal{S}}_q^T}{\partial V}\right)_{\!U_q}\, .
\label{31bb}
\end{eqnarray}
Note that in the limit $q\rightarrow 1$ the conventional relation is recovered.

Let us now see what is a thermodynamic variable conjugate to ${\mathcal{S}}_q^T$.
In fact, Caratheodory theorem~\cite{Caratheodory,CaratheodoryII} ensures that heat one-form has an integration factor, but since the entropy is not additive one cannot use the conventional Carnot cycle argument~\cite{Huang} in the proof of Clausius equality to simply identify the integration factor with the inverse temperature.

Let us dwell a bit more on this last point. 
Let us first assume that heat $Q_q$ is additive in the same way as internal energy [cf. relation~(\ref{23cf})]. We also pass from entropy's natural variables to general state variables 
$\{{\bf{a}}, \vartheta\}$.
Here ${\bf{a}}$ will represent a collection of relevant state variables and $\vartheta$ is the empirical temperature whose existence is guaranteed by the 0th law of thermodynamics [see Eq.~(\ref{29cf})].

Since, by Caratheodory theorem the exact differential associated with the heat one-form is the entropy, one can write
\begin{eqnarray}
d{\mathcal{S}}_q^T({\bf{a}},\vartheta) \ = \ \mu({\bf{a}}, \vartheta) \delta Q_q({\bf{a}},\vartheta)\, ,
\label{32cfg}
\end{eqnarray}
We now divide the analyzed system into two subsystems $A$ and $B$, which are respectively described by state variables
$\{{\bf{a}}_A,\vartheta\}$ and $\{{\bf{a}}_B,\vartheta\}$. Consequently
\begin{eqnarray}
&&\delta Q_q({\bf{a}}_A,\vartheta) \ = \ \frac{1}{\mu_A({\bf{a}}_A,\vartheta)} \ \! d{\mathcal{S}}_q^T({\bf{a}}_A,\vartheta)\, ,\nonumber \\[2mm]
&&\delta Q_q({\bf{a}}_B,\vartheta) \ = \ \frac{1}{\mu_B({\bf{a}}_B,\vartheta)} \ \! d{\mathcal{S}}_q^T({\bf{a}}_B,\vartheta)\, .
\end{eqnarray}
So, for the whole system
\begin{eqnarray}
\delta Q_q(A,B) \ = \ \delta Q_q(A) \ + \ \delta Q_q(B) \, ,
\end{eqnarray}
with 
\begin{widetext}
\begin{eqnarray}
\delta Q_q(A,B)  \ \equiv \ \delta Q_q({\bf{a}}_A,{\bf{a}}_B, \vartheta)
\ = \  \frac{1}{\mu_{A+B}({\bf{a}}_A,{\bf{a}}_B,\vartheta)} \ \! d{\mathcal{S}}_q^T({\bf{a}}_A, {\bf{a}}_B,\vartheta)\, ,~~~~
\end{eqnarray}
and we can write
\begin{eqnarray}
d{\mathcal{S}}_q^T({\bf{a}}_A, {\bf{a}}_B,\vartheta) 
\ = \ \frac{\mu_{A+B}({\bf{a}}_A,{\bf{a}}_B,\vartheta)}{\mu_A({\bf{a}}_A,\vartheta)} \ \! d{\mathcal{S}}_q^T({\bf{a}}_A,\vartheta) \ + \
 \frac{\mu_{A+B}({\bf{a}}_A,{\bf{a}}_B,\vartheta)}{\mu_B({\bf{a}}_B,\vartheta)} \ \! d{\mathcal{S}}_q^T({\bf{a}}_B,\vartheta)\, .
 \label{B.14.fg}
\end{eqnarray}
Let us now assume that there is only one state variable (apart from $\vartheta$), so that ${\bf{a}} = a$. If there were more state variables, our following argument would hold true as well, but we would need to employ more than two subsystems. 
Assuming we can invert ${\mathcal{S}}_q^T({{a}}_A,\vartheta)$ and ${\mathcal{S}}_q^T({{a}}_B,\vartheta)$, we can express $a_A$ and $a_B$ as
\begin{eqnarray}
a_A \ = \ a_A({\mathcal{S}}_q^T(A), \vartheta) \;\;\; \mbox{and} \;\;\; a_B \ = \ a_B({\mathcal{S}}_q^T(B), \vartheta) \, .
\end{eqnarray}
With this, Eq. (\ref{B.14.fg}) can be cast into form
\begin{eqnarray}
d{\mathcal{S}}_q^T({\mathcal{S}}_q^T(A), {\mathcal{S}}_q^T(B),\vartheta)\ &=& \ \frac{\mu_{A+B}({\mathcal{S}}_q^T(A), {\mathcal{S}}_q^T(B),\vartheta)}{\mu_A({\mathcal{S}}_q^T(A),\vartheta)} \ \! d{\mathcal{S}}_q^T(A) 
\ + \
 \frac{\mu_{A+B}({\mathcal{S}}_q^T(A), {\mathcal{S}}_q^T(B),\vartheta)}{\mu_B({\mathcal{S}}_q^T(B),\vartheta)} \ \! d{\mathcal{S}}_q^T(B)\ + \ 0 \ \! d\vartheta\, .~~
 \label{B.16.cb}
\end{eqnarray}
Since $d{\mathcal{S}}_q^T$ (for all considered systems, i.e. $A$, $B$ and $A+B$) must be an exact differential, so that ${\mathcal{S}}_q^T$ is a proper state function, integrability conditions give the following set of equations
\begin{eqnarray}
&&\frac{\partial \log(\mu_A({\mathcal{S}}_q^T(A),\vartheta))}{\partial \vartheta} \ = \ \frac{\partial \log(\mu_B({\mathcal{S}}_q^T(B),\vartheta))}{\partial \vartheta} \ = \ \frac{\partial \log(\mu_{A+B}({\mathcal{S}}_q^T(A), {\mathcal{S}}_q^T(B),\vartheta))}{\partial \vartheta}\, , \label{B.17a.cv} \\[2mm]
&& \frac{1}{\mu_A({\mathcal{S}}_q^T(A),\vartheta)} \frac{\partial \mu_{A+B}({\mathcal{S}}_q^T(A), {\mathcal{S}}_q^T(B),\vartheta)}{\partial {\mathcal{S}}_q^T(B)} \ = \ \frac{1}{\mu_B({\mathcal{S}}_q^T(B),\vartheta)} \frac{\partial \mu_{A+B}({\mathcal{S}}_q^T(A), {\mathcal{S}}_q^T(B),\vartheta)}{\partial {\mathcal{S}}_q^T(A)}\, . \label{B.17b.cv}
\end{eqnarray}
\end{widetext}
Note that the expression (\ref{B.17a.cv}) implies that the derivatives cannot depend on entropy, but only on $\vartheta$. By denoting the RHS of (\ref{B.17a.cv}) as $-w(\vartheta)$,
we might resolve (\ref{B.17a.cv}) in the form
\begin{eqnarray}
&&\mbox{\hspace{-6mm}}\mu_A({\mathcal{S}}_q^T(A),\vartheta) \ = \ \psi_A({\mathcal{S}}_q^T(A)) \ \!\exp\left(-\int w(\vartheta) d\vartheta \right)\, ,\nonumber \\[2mm]
&&\mbox{\hspace{-6mm}}\mu_B({\mathcal{S}}_q^T(B),\vartheta) \ = \ \psi_B({\mathcal{S}}_q^T(B)) \ \!\exp\left(-\int w(\vartheta) d\vartheta \right)\, ,\nonumber \\[2mm]
&&\mbox{\hspace{-6mm}}\mu_{A+B}({\mathcal{S}}_q^T(A),{\mathcal{S}}_q^T(B),\vartheta) \nonumber \\[2mm] 
&&\mbox{\hspace{-1mm}}= \ \psi_{A+B}({\mathcal{S}}_q^T(A),{\mathcal{S}}_q^T(B)) \ \!\exp\left(-\int w(\vartheta) d\vartheta \right),~~
\label{B.19.hh}
\end{eqnarray}
where $\psi$ are some arbitrary functions of the entropy. It is worth stressing $\mu$ factorized into purely entropic and purely temperature-based parts.  In addition, the temperature part of $\mu$ is  not explicitly $q$-dependent.

Let us now observe from (\ref{23cfg}) that
\begin{eqnarray}
&&\mbox{\hspace{-15mm}}d{\mathcal{S}}_q^T({\mathcal{S}}_q^T(A), {\mathcal{S}}_q^T(B),\vartheta) \ = \
d {\mathcal{S}}_q^T(A,B) \nonumber \\[2mm] &&= \ 
\left(1 + \frac{1-q}{k_T}\ \!{\mathcal{S}}_q^T(B) \right) \ \! d{\mathcal{S}}_q^T(A) \nonumber \\[2mm] 
&&+ \ \left(1 + \frac{1-q}{k_T}\ \!{\mathcal{S}}_q^T(A) \right) \ \! d{\mathcal{S}}_q^T(B)\, .
\label{II.20.cf}
\end{eqnarray}
By comparing this equation with (\ref{B.16.cb}) and (\ref{B.19.hh}), we can make the identification $\psi({\mathcal{S}}_q^T) = c \left(1 + \frac{1-q}{k_T}\ \!{\mathcal{S}}_q^T\right) $, with $c$ being an arbitrary constant. At this point, one can easily check that also the second integrability condition
(\ref{B.17b.cv}) is satisfied. In conventional thermodynamics, $\psi$ would be only a constant, and thus the inverse integration factor could be identified with a genuine absolute temperature. In the context of non-additive entropy  ${\mathcal{S}}_q^T$, we see that this is no longer the case. Fortunately, $\mu$ has a simple factorized form, which allows us to rephrase (\ref{32cfg}) as
\begin{eqnarray}
d{\mathcal{S}}_q^T \ = \  \left(1 + \frac{1-q}{k_T}\ \!{\mathcal{S}}_q^T\right) \frac{\delta Q_q}{T_{\rm{phys}}}\, .
\label{43cff}
\end{eqnarray}
Here we, have denoted the temperature part of $\mu$ in (\ref{B.19.hh}) with $1/T_{\rm{phys}}$. Therefore, 
\begin{eqnarray}
T_{\rm{phys}} \ = \ \frac{1}{c} \ \!\exp\left(\int w(\vartheta) d\vartheta \right)\, ,
\end{eqnarray}
plays the role of an absolute temperature in Tsallis thermostatistics. Note that by writing [as in Eqs.~(\ref{29cf}) and  (\ref{31bb})]
\begin{eqnarray}
d{\mathcal{S}}_q^T &=& \left(\frac{\partial {\mathcal{S}}_q^T}{\partial U_q}\right)_{V} dU_q 
\ + \  \left(\frac{\partial {\mathcal{S}}_q^T}{\partial V}\right)_{U_q}dV \nonumber \\[2mm]
&=& \ \left( 1 + \frac{1-q}{k_T} {\mathcal{S}}_q^T  \right) \frac{1}{\vartheta} \ \! dU_q\nonumber \\[2mm] &+& \ \left( 1 + \frac{1-q}{k_T} {\mathcal{S}}_q^T  \right) \frac{p_{\rm{phys}}}{\vartheta} \ \! dV  \, ,
\end{eqnarray}
we obtain 
\begin{eqnarray}
\frac{\vartheta}{T_{\rm{phys}}}\ \!\delta Q_q \ = \ dU_q 
\ + \ p_{\rm{phys}} dV\, .
\label{56.cf}
\end{eqnarray}
The first law of thermodynamics (energy conservation) appears when we identify the empirical temperature introduced in (\ref{29cf}) and  (\ref{31bb}) with $T_{\rm{phys}}$. In turn, this implies $w(\vartheta)= 1/\vartheta$. Additionally, when $T_{\rm{phys}}$ has the same units of the temperature, then $c= 1$. Relations (\ref{43cff}) and (\ref{56.cf}) were first derived in Ref.~\cite{Abe1} using a different approach.

Tsallis (or non-extensive) thermodynamics, which is implied by (\ref{43cff}) and (\ref{56.cf}), is expected to be instrumental in the characterization of statistical systems with long-range interactions~\cite{Abe1} and/or strong correlations~\cite{jizba-korbel:19}. 

Before concluding this section, we point out that  (\ref{43cff}) can be equivalently rewritten as
\begin{eqnarray}
\delta Q_q \ &=& \ \frac{k_T T_{\rm{phys}} }{1-q} \ \! d \ln \left(1 + \frac{1-q}{k_T} \ \! {\mathcal{S}}_q^T  \right)\nonumber \\[2mm]
&=& T_{\rm{phys}} \ \! d {\mathcal{S}}_q^R\, ,
\end{eqnarray}
where ${\mathcal{S}}_q^R$ is the so-called R\'{e}nyi entropy~\cite{Renyi}
\begin{eqnarray}
{\mathcal{S}}_q^R(\mathcal{F}) &=& \frac{k_T}{(1-q)}\ln\left(\int_{\mathbb{R}}d{p}
\,\mathcal{F}^{q}(p)\right)\,,
\end{eqnarray}
that plays an important role in both classical and quantum information theory~\cite{JD:16,JDJ,JDP}. Unless otherwise specified, we will set in the following $k_T=1$.

\section{GUP and Tsallis entropy-power based uncertainty relations}
\label{EUR}

\subsection{Shannon entropy power}
\label{EURG}

By relying on the notion of entropic functional (in this case Tsallis entropic functional) that is extremized by CSs, one can reformulate 
the variance-based GUP (\ref{gup}) in terms of
entropic uncertainty relations. This can be done in line with conventional quantum mechanics~\cite{Hirschman}, where the key aspect is that Glauber CSs  saturate Heisenberg's uncertainty relation (UR) and at the same time maximize Shannon's entropy.
The most compact form of the ensuing entropic UR  can be phrased in terms of (Shannon) {\em entropy power}~\cite{JD:16,JDJ,JDP}. 

Entropy power (EP) was originally introduced by C.~Shannon in order to solve a number of information-theoretic problems related to continuous  random variables~\cite{Shannon}.
%
%
In its essence, EP describes the variance of a would-be Gaussian random variable with the same Shannnon entropy as the random
variable under investigation. Hence, by denoting Shannon's EP of a continuous $D$-dimensional random variable $\mathcal{X}$ as $N(\mathcal{X})$, EP should satisfy the defining relation
\begin{eqnarray}
{\mathcal{S}}^S(\mathcal{X}) \ = \ {\mathcal{S}}^S(\sqrt{N(\mathcal{X})}\cdot
{\mathcal{Z}}^{_G})\, ,
\label{58.cc}
\end{eqnarray}
where $\{{\mathcal{Z}}_{i}^{_G}\}$ represents a $D$-dimensional {\em Gaussian random vector} with zero mean
and unit covariance matrix.
Equation~(\ref{58.cc}) has the unique solution~\cite{Shannon}
\begin{eqnarray}
N(\mathcal{X}) \ = \ \frac{1}{2\pi e}\exp\left(\frac{2}{D} {\mathcal{S}}^S(\mathcal{X})  \right)\, . 
\end{eqnarray}
Conventional canonical commutation relation $[\hat{x},\hat{p}] =i\hbar$ implies that eigenstates of  $\hat{x}$ in {\em momentum representation} or  $\hat{p}$ in {\em position representation} are plane waves. This, in turn, dictates that the $x$ and $p$-representation wave functions $\psi(x)$ and $\hat{\psi}(p)$, respectively, must be related via Fourier transform (here $\hat{\psi}(p)$ should not be mistaken with an operator). If we now  apply the Beckner--Babenko inequality for Fourier-transform duals~\cite{Beckner1975,Babenko1962}, i.e. 
%
%
\begin{eqnarray}
&&\mbox{\hspace{-10mm}}\left[\left(\frac{q'}{2\pi \hbar} \right)^D\right]^{1/q'} |\!| |\psi|^2 |\!|_{q'/2}  \nonumber \\[2mm] 
&&\mbox{\hspace{10mm}}\leq \ \left[\left(\frac{q}{2\pi \hbar} \right)^D\right]^{1/q} |\!| |\hat{\psi}|^2 |\!|_{q/2}\, ,
\label{61sd}
\end{eqnarray}
(where $|\!|X|\!|_{p}$ is the $p$-norm, and $1/q + 1/q'= 1$ with $q\in {\mathbb{R}}^{+}$, i.e., $q'$ and $q$ are H\"{o}lder conjugates), and set $q = q' = 2$,
%
we obtain the entropy power inequality
\begin{eqnarray}
N(|\hat{\psi}|^2)N(|{\psi}|^2)  \  \geq \ \frac{\hbar^2}{4}\, .
\label{61cd}
\end{eqnarray}
It is important to note that this saturates only for CSs~\cite{JDJ,JDP,Hirschman,BB}. Moreover, it can be shown~\cite{JD:16} that (\ref{61cd}) automatsubsumes the variance-based Heisenberg uncertainty relation.
The fact that uncertainty relation (\ref{61cd}) is phrased entirely in terms of Shannon's entropy is, in part, a reason why Shannon's entropy plays such a key role in conventional quantum mechanics and quantum information theory.

\vspace{3mm}
\subsection{Entropy powers based on Tsallis distribution}
\label{EURT}

When dealing with GUP saturated by Tsallis probability amplitude states, it is convenient to work with EP based on
Tsallis distribution {(\ref{15aa})}.  By emulating the procedure outlined in the previous subsection, we define the entropy powers associated with Tsallis entropies  as solutions of the equation
\begin{eqnarray}
{\mathcal{S}}_q^T
\left( {\mathcal{X}} \right)
\ = \ {\mathcal{S}}_q^T\left(\sqrt{M_q^T(\mathcal{X})}\cdot
{\mathcal{Z}}^{T}\right)\, ,
\label{62kl}
\end{eqnarray}
where $\{{\mathcal{Z}}_{i}^{T}\}$ represents a {\em Tsallis random vector} with zero mean
and unit covariance matrix. Such a vector is distributed with respect to the $q$-Gaussian probability density function that extremizes ${\mathcal{S}}_q^T$.
To solve Eq.~(\ref{62kl}), we first use the scaling relation for Tsallis entropy, namely
\begin{eqnarray}
{\mathcal{S}}_q^T(a \mathcal{X}) \ &=& \ {\mathcal{S}}_q^T(\mathcal{X}) \ \oplus_q \ \ln_q |a|^D\, ,
\label{63cf}
\end{eqnarray}
where $a \in \mathbb{R}$ and the $q$-deformed logarithm (or simply $q$-logarithm) is defined as in~\cite{Tsallis2}
\begin{equation}
\ln_q x  \,=\,  \frac{x^{1-q }- 1}{1-q}\,.
\end{equation}
The relation (\ref{63cf}) follows directly from the chain of identities
\begin{eqnarray}
&&\hspace{-4mm}{\mathcal{S}}_q^T(a \mathcal{X})\nonumber \\[2mm]
&&=  \ \frac{1}{1-q} \left[\int d^D {\bf y}\left(\int d^D {\bf x} \ \! \delta({\bf y} - a {\bf x}) {\mathcal{F}}({\bf{x}}) \right)^{\!q} -1 \right] \nonumber \\[2mm]
&&= \ |a|^{D(1-q)}\ \!{\mathcal{S}}_q^T(\mathcal{X}) \ + \ \ln_q |a|^D
\nonumber \\[2mm]
&&= \ {\mathcal{S}}_q^T(\mathcal{X}) \ \oplus_q \ \ln_q |a|^D\, .
\label{scaling}
\end{eqnarray}
In the second step, we make use of 
\begin{widetext}
\begin{eqnarray}
{\mathcal{S}}_q^T\left({\mathcal{Z}}^{T}\right) \ &=& \ \ln_q \left[\left(\frac{\pi}{\mathfrak{b}(1-q)}  \right)^{\frac{D}{2}} \ \! \frac{\Gamma\left(\frac{1}{1-q}-\frac{D}{2} \right)}{\Gamma\left(\frac{1}{1-q} \right)} \ \!  \left(1- \frac{D}{2q}(1-q)\right)^{\frac{1}{(q-1)}} \right]\, ,
\label{15.bc}
\end{eqnarray}
with  $\mathfrak{b}  =   [2q - D(1-q)]^{-1}$.
%
%
Combining (\ref{62kl}), (\ref{63cf}) and (\ref{15.bc}), we arrive at the Tsallis EP
\begin{eqnarray}
M_q^T(\mathcal{X}) &=& A_q \ \! \left[\exp_q({\mathcal{S}}_q^T(\mathcal{X}))\right]^{2/D} \ = \  A_q \ \! \exp_{1-(1-q)D/2}\left({\frac{2}{D} \ \!{\mathcal{S}}_q^T(\mathcal{X})}\right)\, ,
\label{SM.61cc}
\end{eqnarray}
where the constant $A_q$ is defined as
\begin{equation}
A_q \ = \ \left[\left(\frac{\pi}{\mathfrak{b}(1-q)}  \right)^{\frac{D}{2}} \ \! \frac{\Gamma\left(\frac{1}{1-q}-\frac{D}{2} \right)}{\Gamma\left(\frac{1}{1-q} \right)} \ \!  \left(1- \frac{D}{2q}(1-q)\right)^{\frac{1}{(q-1)}}  \right]^{-{2}/{D}}.
\end{equation}
\end{widetext}
In the above derivation, we employed the sum rule for the $q$-deformed calculus: $\ln_q x \oplus_q  \ln_q y = \ln_q xy$, as well as the definition of the $q$-exponential  $e_q^x = [1 + (1-q)x]^{1/(1-q)}$ and the fact that $e_q^{\ln_q x} = \ln_q (e_q^x) = x$.
%
%

As a consistency check, we might notice that in the $q \rightarrow 1$ limit Shannon's EP is recovered:
\begin{eqnarray}
\lim_{q\rightarrow 1 }M_q^T(\mathcal{X})  
 \ = \ \frac{1}{2\pi e} \exp\left( \frac{2}{D} {\mathcal{S}}^S(\mathcal{X})\right) \ = \ \ N(\mathcal{X})\, .~~
\end{eqnarray}



\subsection{Entropy power inequalities for GUP transformations }
\label{EPIGUP}

By analogy with the conventional commutation relations, where the ensuing entropic URs are of the form~(\ref{61cd}), one can derive entropic URs also for the DCR~(\ref{comm}).  In this latter case it can be expected that the role of Shannon's entropy will be overtaken by Tsallis' entropy (\ref{30cf}).  The actual rationale behind this fact is not difficult to understand. First, we notice (cf. Appendix~A) that for $\beta<0$ the eigenstate of the position operator in the momentum representation is
\begin{eqnarray}
\psi_x(p) \, = \, B_x \ \!
\frac{e^{-ixm_p{\mbox{\scriptsize{arctanh}}}\left(p \sqrt{|\beta|}/m_p  \right)/\hbar\sqrt{|\beta|}}}{\sqrt{m^2_p  \, - \, p^2 |\beta|}}\, ,
\label{70aab}
\end{eqnarray}
with $B_x = \sqrt{m_p^2/2\pi \hbar}$. This implies that the position and momentum representations of a wave function are related not via conventional Fourier transform but, instead, via {\em Abel transform} (cf. Appendix~\ref{negbetaConn})
\begin{eqnarray}
&&\mbox{\hspace{-10.5mm}}\psi(x) \nonumber \\[2mm]
&&\mbox{\hspace{-10.5mm}}=   \int_{\frac{-m_p}{\sqrt{|\beta|}}}^{\frac{m_p}{\sqrt{|\beta|}}} \frac{dp}{\sqrt{2\pi \hbar}} \ \! \frac{e^{ixm_p{\mbox{\scriptsize{arctanh}}}\left(p \sqrt{|\beta|}/m_p  \right)/\hbar\sqrt{|\beta|}}}{\sqrt{1   -  p^2 |\beta|/m_p^2}}\ \! \tilde{\psi}(p)\,,
\label{71.kk}
\end{eqnarray}
(to distinguish it from the HUP-based QM,  the conjugate wave function of $\psi(x)$ is denoted as $\tilde{\psi}(p)$ here).
This does not allow to use the Beckner--Babenko inequality directly, but we might observe that after the substitution 
\begin{eqnarray}
 z \ = \   m_p \ \! \mbox{{arctanh}}\left(p \sqrt{|\beta|}/m_p \right)/\sqrt{|\beta|}\, ,
\end{eqnarray} 
we can rewrite (\ref{71.kk}) as
\begin{eqnarray}
\mbox{\hspace{-10.5mm}}\psi(x) 
\ = \ \int_{\mathbb{R}} \frac{dz}{\sqrt{2\pi \hbar}} \ \! e^{ixz/\hbar} \bar{\psi}(z)   \, ,
\label{72.cdd}
\end{eqnarray}
where 
\begin{equation}
   \bar{\psi}(z) \ = \  \frac{\tilde{\psi}(m_p\tanh(z \sqrt{|\beta|}/m_p)/\sqrt{|\beta|})}{ {\cosh(z \sqrt{|\beta|}/m_p)} }.
   \label{72.cf}
\end{equation}
This allows the use the Beckner--Babenko inequality in the form 
\begin{eqnarray}
 &&\mbox{\hspace{-0.8cm}}\left[\left(\frac{q'}{2\pi \hbar} \right)^D\right]^{1/q} |\!| |\bar{\psi}|^2 |\!|_{q'/2} \nonumber \\[2mm]
 &&\mbox{\hspace{1.3cm}}\leq\ \left[\left(\frac{q}{2\pi \hbar} \right)^D\right]^{1/q'} |\!| |\psi|^2 |\!|_{q/2}
\, ,
\label{45ccc}
\end{eqnarray}
where $q'$ and $q$ are H\"{o}lder conjugates with $q' \in [2,\infty)$, so that $q \in [1,2]$.
It can be checked numerically that (\ref{45ccc}) is saturated by the CSs $\tilde{\psi}_{CS}(p)$ with the non-extensivity index $2-q'/2$ (i.e., Tsallis distributions 
$|\tilde{\psi}|^2_{CS}(p)=
q_T(p|2-q'/2,b)$)  and associated
$\psi_{CS}(x)$  with the non-extensivity index $2-q/2$ [see Eq.~(\ref{CS-x})]. The analytical proof can be readily done for the cases $q=1$ and $q=2$.

Now, we can follow the same steps as in the case of Shannon EPURs~\cite{JLLP}. After some algebra we arrive at
\begin{eqnarray}
M_{q/2}^T(|{\psi}|^2)M_{q'/2}^T(|\tilde{\psi}|^2) &\geq&  
 \frac{\hbar^2}{4} \frac{q^2}{(3q/2-1)(3q'/2-1)} \nonumber \\[2mm]
 &=&  \frac{\hbar^2}{4} f(q)\, .
\label{76.cd}
\end{eqnarray}
The function $f(q)$ is positive and monotonically increasing for $q\in [1,2]$ with $\max f(q) = 1$.
It is important to stress that $f(q)$ depends only on $q$, while no other GUP parameters are present.
We may now use the identity
\begin{eqnarray}
\nonumber
f(q) &=&  \left[\frac{2}{(|2/q -1| + 1) (3q/2-1)}\right]\\[2mm]
\nonumber
&&\times\left[\frac{2}{(|2/q' -1| + 1)(3q'/2-1)}\right] \\[2mm]
&=&\phi(q/2)\phi(q'/2)\, ,
\end{eqnarray}
and rewrite the EPUR (\ref{76.cd}) in the form
\begin{eqnarray}
\tilde{M}_{q/2}^T(|{\psi}|^2)\tilde{M}_{q'/2}^T(|\tilde{\psi}|^2) \  \geq  \ \frac{\hbar^2}{4}\, .
\label{78cf}
\end{eqnarray}
where we have defined the ``rescaled entropy power'' $\tilde{M}_{x}^T = \phi^{-1}(x)M_{x}^T$. This EPUR clearly emulates the form of Shannon's EPUR (\ref{61cd}) by having the irreducible universal lower bound.

Since  the Tsallis
entropy ${\mathcal{S}}_{q'/2}^T$ 
is maximized by the Tsallis distribution
with the non-extensivity parameter $2- q'/2$, i.e. by $q_T(p|2- q'/2,b)$ and because for $\beta < 0$ the non-extensivity parameter
$2- q'/2 < 1$, we have the conditions $q' > 2$ and  $q < 2$. Note that both these conditions are also required by Beckner--Babenko inequality, cf.~Ref~\cite{JLLP}.  The formulation of uncertainty relation (\ref{78cf}) in terms of Tsallis' entropy suggests that Tsallis' entropy should be essential to GUP quantum mechanics and its associated quantum information theory,  much in the same way that Shannon's entropy is essential to conventional quantum mechanics.


\section{Cosmological and gravitational  applications}
\label{CosmApp}

\subsection{Generalization of DCR to 3 dimensions and associated CS}
\label{3D generalization}

In order to discuss prospective cosmological implications, it is important to generalize our 1-dimensional formalism to 3 dimensions. This can be done, for instance, by extending the algebra~(\ref{1.cf})-(\ref{2.cf}) as (see e.g. Refs.~\cite{Kempf,Tawfik})
\begin{eqnarray}
&&\left[\hat{\bf{x}}_i, \hat{\bf{p}}_j \right] \ = \ i \hbar \delta_{ij} \left(1 \ + \ \beta \frac{{\bf{p}}^2}{m_p^2}  \right)\, , \nonumber \\[2mm]
&& \left[\hat{\bf{p}}_i, \hat{\bf{p}}_j \right] \ = \ 0\, , 
\label{79.aa}
\end{eqnarray}
where the ensuing commutator $\left[\hat{\bf{x}}_i, \hat{\bf{x}}_j \right] $ will be chosen after the explicit representation of $\hat{\bf{x}}_i$ is deduced from (\ref{79.aa}). In particular, in the momentum
space,  $\hat{\bf{x}}_i$  and  $\hat{\bf{p}}_i$ satisfying DCR (\ref{79.aa}) can be represented by
[cf. Eqs.~(\ref{operators})]
\begin{eqnarray}
&&\mbox{\hspace{-7mm}}\hat{\bf{p}}_i\,\psi({\bf{p}})\ = \ {\bf{p}}_i\,\psi({\bf{p}})\,, \nonumber \\[2mm]
&&\mbox{\hspace{-7mm}}\hat{\bf{x}}_i\,\psi({\bf{p}})\ = \ i\hbar\lf(\frac{d}{d {\bf{p}}_i}+\frac{\beta}{2\,m_p^2}\ \! \left\{ {\bf{p}}^2,\frac{d}{d {\bf{p}}_i}\right\}\ri)\psi({\bf{p}})\,,
\label{operatorsIII}
\end{eqnarray}
thus implying that
\begin{eqnarray}
\left[\hat{\bf{x}}_i, \hat{\bf{x}}_j \right] \ = \ 2 i \hbar \frac{\beta}{m_p^2} \left( \hat{\bf{x}}_i \hat{\bf{p}}_j - \hat{\bf{x}}_j \hat{\bf{p}}_i\right)\, .
\label{81.cd}
\end{eqnarray}
As a consistency check, one can verify that the commutators (\ref{79.aa}) and (\ref{81.cd}) satisfy the Jacobi identity. 

By analogy with Section~\ref{CSGUP} we can deduce corresponding URs. These read
\begin{eqnarray}
&&(\Delta{\bf{x}}_i)_{\varrho} (\Delta{\bf{p}}_j)_{\varrho} \ \geq \  \frac{\hbar}{2}\delta_{ij}\lf(1+\beta\,\frac{(\Delta{\bf{p}})^2_{\varrho} + \langle \hat{\bf{p}} \rangle^2_{\varrho}  }{m_p^2}\ri)\, , \nonumber \\[2mm]
&& (\Delta{\bf{x}}_i)_{\varrho} (\Delta{\bf{x}}_j)_{\varrho} \ \geq \ 
\hbar \frac{\beta}{m_p^2} \left|\left\langle \left( \hat{\bf{x}}_i \hat{\bf{p}}_j - \hat{\bf{x}}_j \hat{\bf{p}}_i\right)\right \rangle_{\varrho}\right|
\, , \nonumber \\[2mm]
&& (\Delta{\bf{p}}_i)_{\varrho} (\Delta{\bf{p}}_j)_{\varrho} \ \geq \ 0\, .
\label{splitII}
\end{eqnarray}
In the following we will again consider only a mirror symmetric density matrix $\varrho$ satisfying
$\langle \hat{\bf{p}} \rangle_{\varrho} = 0$, so as to attain the GUP~(\ref{gup}).
For simplicity, we will concentrate only on isotropic density matrices, for which
$\langle \hat{\bf{x}} \rangle_{\varrho} = 0$. This will be fully satisfactory for our subsequent reasoning.

As will be shown shortly, on the class of mirror symmetric $\varrho$'s, the equation [cf. Eq.~(\ref{eqn})]
%
%
%
%
\be\label{eqn.83.nm}
\lf(\hat{\bf{p}}_k-i\gamma_k \hat{\bf{x}}_k\ri)|\psi\rangle \ = \ 0\,, \;\;\; k = 1,2,3\, ,
\ee
admits only one solution for $\psi \in L^2(\mathbb{R})$, so that the minimum–uncertainty $\hat{\varrho}$  is a pure (coherent) state.
To see this, let us first introduce the generator of rotations\footnote{It is worth observing that, in the context of the GUP, the definition of the angular momentum might not coincide with the generator of the rotations. For further details on this  aspect, the interested reader can consult Refs. \cite{angmom,angmom2}.}. Bearing in mind the DCR (\ref{79.aa}) and (\ref{81.cd}), we can define~\cite{Kempf}
\begin{eqnarray}
\hat{\ \!\bf{L}}_k \ = \ \frac{1}{1+ \frac{\beta}{m_p^2}\hat{\bf{p}}^2} \ \! \epsilon_{klm} \ \!\hat{\bf{x}}_l \hat{\bf{p}}_m\, ,
\end{eqnarray}
where $\epsilon_{klm}$ is the Levi-Civita antisymmetric symbol. These operators satisfy the standard \emph{so}(3) algebra, i.e.
\begin{eqnarray}
\left[\hat{\ \!\bf{L}}_k, \hat{\ \!\bf{L}}_l\right] \ = \ i\hbar \ \! \epsilon_{klm} \ \! \hat{\ \!\bf{L}}_m\, , 
\end{eqnarray}
alongside with the other familiar commutation relations
\begin{eqnarray}
&&\left[\hat{\bf{x}}_k, \hat{\ \!\bf{L}}_l\right] \ = \ i\hbar \ \! \epsilon_{klm} \ \! \hat{\bf{x}}_m\, ,
\nonumber \\[2mm]
&&\left[\hat{\bf{p}}_k, \hat{\ \!\bf{L}}_l\right] \ = \ i\hbar \ \! \epsilon_{klm} \ \! \hat{\bf{p}}_m\, .
\end{eqnarray}
With the operator $\hat{\ \!\bf{L}}_k$ at hand, we rewrite $(\Delta{\bf{x}}_i)_{\varrho} (\Delta{\bf{x}}_j)_{\varrho}$ in (\ref{splitII}) in the form
\begin{eqnarray}
(\Delta{\bf{x}}_i)_{\varrho} (\Delta{\bf{x}}_j)_{\varrho}\ &\geq& 
\frac{\beta}{m_p^2} \left| \left\langle \left(1 + \frac{\beta}{m_p^2} {\hat{\bf{p}}}^2  \right) \! \left[\hat{\ \!\bf{L}}_i, \hat{\ \!\bf{L}}_j \right] \right\rangle_{\!\!\varrho}\ \!\right|
\nonumber\\[2mm]
&&\mbox{\hspace{-15mm}}=\ \frac{\beta}{m_p^2} \left|\mbox{Tr}\!\left[\left(1 + \frac{\beta}{m_p^2} {\hat{\bf{p}}}^2  \right) \!\hat{\ \!\bf{L}}_i\left[\hat{\ \!\bf{L}}_j, \hat{\varrho} \right]\right]\right|
\nonumber\\[2mm]
&&\mbox{\hspace{-15mm}}
=\ \frac{\beta}{m_p^2} \left|\mbox{Tr}\!\left[\left(1 + \frac{\beta}{m_p^2} {\hat{\bf{p}}}^2  \right) \!\hat{\ \!\bf{L}}_j\left[\hat{\ \!\bf{L}}_i, \hat{\varrho} \right]\right]\right|
\nonumber\\[2mm]
&&\mbox{\hspace{-15mm}}=\ 0
\, .
\label{87.cdv}
\end{eqnarray}
The last line is a consequence of the assumed isotropy of  $\hat{\varrho}$.

To find the CS, we return back to  Eq.~(\ref{eqn.83.nm}). Because the isotropy of  $\hat{\varrho}$ implies $\hat{\bf{L}}_j| \psi \rangle = 0$ for $j=1,2,3$, we obtain from~(\ref{eqn.83.nm}) that 
\begin{eqnarray}
0 &=& \hat{\bf{L}}_j \lf(\hat{\bf{p}}_k-i\gamma_k \hat{\bf{x}}_k\ri)|\psi\rangle 
\ \! = \ \! \left([\hat{\bf{L}}_j,\hat{\bf{p}}_k] - i\gamma_k [\hat{\bf{L}}_j,\hat{\bf{x}}_k]   \right) |\psi\rangle \nonumber \\[2mm]
&=& -i\hbar \varepsilon_{jkl} \lf(\hat{\bf{p}}_l-i\gamma_k \hat{\bf{x}}_l\ri)|\psi\rangle \, .
\end{eqnarray}
This, for instance, gives
\begin{eqnarray}
\gamma_1 \hat{\bf{x}}_l|\psi\rangle  &=&  \gamma_2 \hat{\bf{x}}_l|\psi\rangle \nonumber  \ = \  \gamma_3 \hat{\bf{x}}_l|\psi\rangle\, ,
\end{eqnarray}
for any $l = 1, 2, 3$, which ensures that $\gamma_k$ is $k-$independent ($|\psi\rangle$'s satisfying (\ref{eqn.83.nm}) cannot be all zero-eigenvalue states of $\hat{\bf{x}}_l$ as $(\Delta{\bf{x}}_l)_{\varrho} \neq 0$). With this, Eq.~(\ref{eqn.83.nm})
can be rewritten in the form
\begin{eqnarray}
\frac{\partial}{\partial {\bf{p}}_k}\,\psi({\bf{p}}) \ = \ -\frac{\lf(1+\frac{\beta\ga\hbar}{m_p^2}\ri)}{\ga\hbar\lf(1  +  \frac{\beta}{m_p^2}{\bf{p}}^2\ri)}\,{\bf{p}}_k\,\psi({\bf{p}})\,,
\end{eqnarray}
This system of 3 equations admits only one (normalized) solution for $\psi \in L^2(\mathbb{R}^3)$, namely
\begin{eqnarray}
\psi({\bf{p}})  \ = \ N\left[1  \  + \   ({\beta\,{\bf {p}}^2})/{m_p^2}\right]_{+}^{-\frac{m_p^2}{2\beta\ga\hbar}-\frac{1}{2}}\, ,
\label{90.cc}
\end{eqnarray}
%
with the normalization constant
\begin{eqnarray}
&&\mbox{\hspace{-6mm}}N_{_>}^2 \ = \ \frac{\beta^{3/2}}{2\pi m_p^3 \ \!B(5/2, {m_p^2}/{\beta \gamma\hbar} -3/2)}\, ,\;\; {\mbox{for}} \;\;\ \beta > 0\, , \nonumber \\[2mm]
&&\mbox{\hspace{-6mm}}N_{_<}^2 \ = \  \frac{\beta^{3/2}}{2\pi m_p^3 \ \!B(5/2, {m_p^2}/{|\beta| \gamma\hbar} )}\, ,\;\; {\mbox{for}} \;\;\ \beta < 0\, ,
\end{eqnarray}
where $B(x,y)$ is the beta function.
The uniqueness of the solution (\ref{90.cc}) ensures that the minimum–uncertainty $\hat{{\varrho}}$ is again a pure CS.

Note that the CS (\ref{90.cc}) is indeed as close as we can get to the classical situation. Namely, the irreducible non-zero lower bound is saturated and all other uncertainty relations are bigger than zero, which is also true in classical physics.  

Among all pointer states in the would-be
GUP driven Universe, only CSs~(\ref{90.cc}) (similarly to their 1D counterparts~(\ref{sol}))
saturate both the ``${\bf{x}}$-${\bf{p}}$''~GUPs and ensuing Tsallis EPURs.
Moreover, in Section~\ref{EPIGUP} we have seen that the very existence of
Tsallis EPUR indicates that TE should be a relevant entropy functional in the GUP context.
If we couple this observation with the fact that 
CSs~(\ref{90.cc}) extremize Tsallis entropy in 3D space, we might invoke (similarly as in conventional statistical physics) MEP but this time with Tsallis entropy (in place
of Shannon--Gibbs entropy) 
to discuss a statistical physics of an ensemble of non-interacting GUP-governed particles 
in their quasi-classical regime.
In this respect, {\em non-extensive thermodynamics of Tsallis}~\cite{Tsallis1,Tsallis2} provides the necessary mathematical framework that can be utilized to explore the quasi-classical domain of a GUP Universe. 


\subsection{Verlinde's entropic gravity}
\label{Verlinde}



In Ref.~\cite{JLLP} we have used the entropy one-form (\ref{43cff}) to show that Verlinde's entropic-gravity force~\cite{Verlinde} defined by the relation 
$F\delta x  \ = \ T \delta S$, 
($S$ is the holographic entropy obeying the area-law scaling --- basically the Bekenstein--Hawking entropy)  generalizes in the present context and yields the gravitational potential
\begin{eqnarray}
V(r) \ = \ \frac{r_s}{2} \left[ -\frac{1}{r} \ + \ (1-q) \kappa_2 \ \!r\right]\, , 
\label{92.cf}
\end{eqnarray}
where $\kappa_2 = \omega_2/\ell_p^2$ ($\ell_p = \hbar/m_p \approx 1.6 \times 10^{-35}$m is the Planck length, $\omega_2 = \pi$ is the second Hill's coefficient~\cite{JLLP,Hill}) and $r_s = 2M$ is the Schwarzchild radius (here and in subsequent reasonings we adopt the convention that the gravitational constant $G=1$).  The gravitational potential (\ref{92.cf}) coincides with the Mannheim--Kazanas gravitational potential of a static, spherically symmetric source
of mass $M$ in conformal Weyl gravity (CWG)~\cite{mannheim_a,mannheim_b, mannheim_c}. There, a parameter in front
of the linear term 
is identified with the inverse Hubble length $R_H$ (more precisely with $1/(2R_H)$)~\cite{Kazanas}.
What is quite intriguing here is that,
for present macroscopic scales (i.e., $R_H \sim 10^{26}$m), the Mannheim--Kazanas solution has been successful in fitting more than two hundred galactic rotation curves with no adjustable parameters (other than the galactic mass-to-light ratios) and with no need for dark matter or other exotic modifications of gravity~\cite{mannheim_a,mannheim_b, mannheim_c}. 
Despite the fact that macroscopic-scale gravity does not fall within the assumed quasi-classical regime, the idea that the coefficient in front of a linear term in (\ref{92.cf}) should be associated with the inverse Hubble length is valid even in the early Universe cosmology. This is because the argument of CWG
leading to this result is independent of an actual Universe epoch~\cite{MK2}. 
%
%

In conventional cosmology, it is expected that a quasi-classical (decoherence) description  becomes pertinent at the late-inflation epoch (after the first Hubble radius crossing) and perhaps even after its end during reheating~\cite{KieferII,Burges}. 
So, in this period the NTT should be a suitable framework for the description of an ``inflaton gas''.  For instance, by viewing the ``inflaton gas'' as an ideal gas,
the NTT predicts that the inflaton pressure should satisfy for $0< q < 1$ a polytrope relation~\cite{Du:04,Abe4}
\begin{eqnarray}
p \ = \ \frac{2\pi \hbar^2}{m_i\ \! e^{5/3}} \ \! \rho^{5/3}\, ,
\label{93.cf}
\end{eqnarray}
where $\rho = N/V$ is the particle density and $m_i$ the
mass of the inflaton. 
In this connection, it should be stressed that the relation (\ref{93.cf}) holds for $0< q < 1$ but not in the limit $q \rightarrow 1$, see, e.g.~\cite{Abe4}. In fact, at $q=1$ one has the familiar pressure relation $p \propto \rho_E$. So, the
NTT and extensive limits are not interchangeable.
The polytrope relation of the type (\ref{93.cf}) often appears in phenomenological studies on late inflation, see, e.g.~\cite{Bar:16,Bar:18}.


In order to gain information about $\beta$, we employ the CWG observation that cosmologically viable linear term in (\ref{92.cf}) should have its parameter  associated with $1/R_H$.
According to CWG 
the Newtonian potential 
(\ref{92.cf})
should dominate on short scales, while  the linear one becomes prominent at large scales. Both potentials get equal at $R_H$, which in our case implies that $q= 1- \ell_p^2/(\pi R_H^2)$.   Note that this is compatible with the condition that $r_s = R_H$.
%
By 
combining the latter expression for $q$ with  ~\eqref{13cd} and (\ref{17.ccd}),
we obtain $|\beta| \simeq m_p^2\, \ell_p^2/(2\pi\left(\Delta p\right)^2_\psi R_H^2)$. 

To see how such $\beta$ explicitly depends on a cosmological time $t$, we 
first write $R_H(t)=H^{-1}(t)=a(t)/\dot{a}(t)$,
where $H$ is the Hubble parameter
and $a(t)$ is the scale factor. The latter can be evaluated, e.g., from the Vilenkin--Ford inflationary model~\cite{VF}, where
$a(t)=A\sqrt{\mathrm{sinh}\left(B\,t\right)}$, with $B=2\sqrt{\Lambda/3}$
($\Lambda$ is the cosmological constant).
We then use the relativistic equipartition theorem 
$\left(\Delta p\right)^2_\psi \simeq 12 \left(k_B T\right)^2$, cf.~Ref~\cite{JLLP}. 
After simple algebraic manipulations, we obtain
\begin{eqnarray}
\label{betaest2}
|\beta| \ \equiv \ |\beta(t)|\ = \ \frac{m_p^2\, \ell_p^2\,\Lambda}{72\pi\left(k_B T\right)^2\mathrm{tanh}^2\left({2t\sqrt{{\Lambda}/{3}}}\right)}\, .
\end{eqnarray}
%
%
For concreteness' sake, let us consider the reheating  epoch,
i.e. time scale $t\simeq 10^{-33}$s. By taking $m_i = 10^{12}\div 10^{13}$GeV,
$T$ of the order of the reheating temperature $T_R\simeq 10^{7} \div 10^8$GeV and 
the presently known value of
the cosmological constant 
$\Lambda \simeq 10^{-52}$m$^{-2}$,
we obtain $|\beta| \sim  10^{-2} \div 1$, which is in agreement with the values predicted by string-theory 
models, cf. e.g.~\cite{VenezGrossMende,Ciafa,Koni}. This result is also consistent with the naturalness principle that  dictates that not so far from the Planck scale the $\beta$ should not be too large nor too small. 
Let us finally re-emphasize that the above connection with the CWG exists only when $\beta<0$, as otherwise 
the linear term in (\ref{92.cf})
would have an erroneous sign.

%
%

\subsection{Loop Quantum Gravity}
Loop Quantum Gravity (LQG) is a non-perturbative and background-independent theory of quantum gravity, characterized by quantum operators
for areas and volumes that exhibit discrete spectra~\cite{Rovelli}. 
A basic postulate is that the spacetime structure is formed  by finite loops nested into 
extremely fine networks - the
spin networks. These are graphs
with edges having labels $j=0,1/2,1,3/2\dots$. As shown in~\cite{Rovelli2}, the area element
carried by a given surface punctured 
by the spin network edge $j$ is
\begin{equation}
\label{aj}
a(j)=8\pi\ell_p^2\gamma_{LQG}\sqrt{j(j+1)}\,,
\end{equation}
where $\gamma_{LQG}$ is the Immirzi parameter~\cite{Immirzi}, which is a positive, real-valued number that measures the size of a quantum of area in Planck units.

\subsubsection{Black hole quasinormal modes}

While providing a fundamental prediction of LQG, Eq.~\eqref{aj}
is beset by the ambiguity that $\gamma_{LQG}$ is in principle undetermined. A possible way to fix it is by connecting the relation between the area and mass of a Schwarzschild black hole to the area produced by the spin network through the definition of quasinormal modes~\cite{QNM}, which are a set of damped oscillations satisfying the perturbation equations of the Schwarzschild geometry~\cite{ReW}.
In so doing, the minimum value $j_{min}=1$ has been obtained in~\cite{QNM}, which in turn fixes the Immirzi parameter to 
\begin{equation}
\label{standgamma}
\gamma_{LQG}=\frac{\log3}{\pi\sqrt8}\,.
\end{equation}
Also, we emphasize that the result
$j_{min}=1$ has been interpreted  in~\cite{QNM} by assuming that the effective gauge group of the spin networks to consider in LQG is SO(3) (whose unitary representations are in fact labeled by integers) instead of its covering (and normally adopted) group SU(2). 

The above considerations apply to the
case where the standard Boltzmann-Gibbs statistics is used for black holes. However, non-trivial conclusions on the gauge structure of LQG can be reached within Tsallis' framework, as recently suggested in~\cite{AbreuQN}.
To show how non-extensivity - and, consequently, the GUP - affects black holes quasinormal modes, let us recall that quasinormal
mode frequencies $\omega_n$ for large damping are limited by~\cite{Damp}
\be
M\omega_n=0.04371235+\frac{i}{4}\left(n+\frac{1}{2}\right),
\ee
with $M$ and $n$ being the mass of the black hole and a non-negative integer, respectively. Moreover, the real part $w_n\equiv Re[\omega_n]$ of these frequencies obeys
$w_n=\log 3/(8\pi M)$~\cite{Hod}.

Based on~\cite{Hod}, one can assume that the variation $\Delta M$ in the mass of the black hole 
equals the energy of a quantum with frequency $w_n$, i.e. $
    \Delta M=\hbar w_n=\hbar \log3/(8\pi M)$.
Combining this equation with the standard mass/area relation $A=16\pi M^2$ for a Schwarzschild black hole, we get the corresponding surface change $\Delta A=4\hbar\log3$.
This result can be used to relate the Immirzi parameter to the value of $j_{min}$. According to Bohr's correspondence principle, the oscillatory frequency 
of a classical system should correspond to the transition frequency of the analogue quantum system. In the framework of LQG, the most natural description of a black hole ``transition'' is in terms of 
the appearance or disappearance
of a puncture with spin $j_{min}$.
The ensuing change in the area of the black hole can be quantified by Eq.~\eqref{aj} with $j=j_{min}$, yielding in the end
\begin{equation}
\gamma_{LQG}=\frac{\log 3}{2\pi\sqrt{j_{min}(j_{min}+1)}}\,.
\label{gLQG}
\end{equation}

To infer the dependence of $j_{min}$ in Tsallis statistics, we should first evaluate the number of configurations in a punctured surface, taking into account the multiplicity of each state $j$. By resorting to Tsallis entropy, the following expression is then obtained for the case of a microcanonical ensemble~\cite{AbreuQN}
\begin{equation}
\label{jmin}
j_{min}=\frac{1}{2}\left\{\left[1+(1-q)\frac{A}{4\ell_p^2}\right]^{\frac{\log 3}{(1-q)\frac{A}{4\ell_p^2}}}-1\right\}.
\end{equation}
Thus, non-extensivity introduces an extra degree of freedom, resulting into a different spectrum for $j_{min}$. Clearly, 
the standard result $j_{min}=1$ is recovered in the $q\rightarrow1$ limit, which in turn gives the  value~\eqref{standgamma} of $\gamma_{LQG}$.
Also, Eq.~\eqref{jmin} enables to constrain departure from extensivity $1-q$. Toward this end, we notice that the lowest (nonzero) allowed spin $j_{min}=1/2$ is obtained, provided that $1-q\approx1.37\,\frac{4\ell_p^2}{A}$, while $j_{min}\rightarrow\infty$ for $1-q\rightarrow-4\ell_p^2/A$.

\begin{figure}[t]
\begin{center}
\includegraphics[width=8.3cm]{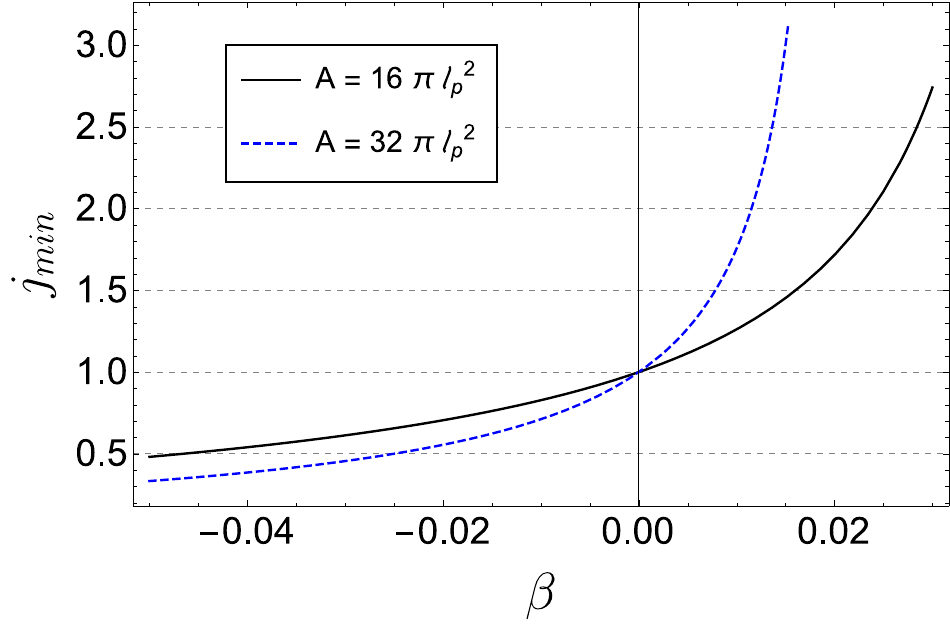}
\caption{Values of $j_{min}$ as a function of $\beta$ for  $A=16\pi\ell_p^2$ (black solid line) and $A=32\pi\ell_p^2$ (blue dashed line). We set the energy scale to $\Delta p\simeq m_p$ (online colors).}
\label{miva}
\end{center}
\end{figure}

Using Eq.~\eqref{qb}, we can now convert the $q$-dependence of $j_{min}$ into a $\beta$-dependence to see how GUP interfaces with LQG. After some algebra, we are led to
\begin{eqnarray}
 &&\mbox{\hspace{-4mm}}j^{(\beta)}_{min} \ =\nonumber \\[2mm]
 &&\hspace{-4mm}\frac{1}{2}\left\{\left[\frac{1+\beta \Delta p^2\left(\frac{3}{m_p^2}-\frac{A}{2\hslash^2}\right)}{1+3\beta\frac{\Delta p^2}{m_p^2}}\right]^{-\frac{2\hslash^2\log 3}{A}\left(\frac{3}{m_p^2}+\frac{1}{\beta\Delta p^2}\right)}-1
    \right\}.\nonumber \\[2mm]
    \label{jminb}
\end{eqnarray}
%
In Fig.~\ref{miva} we plot this expression as a function of $\beta$ for a micro black hole of area $A=16\pi\ell_p^2$ and for $A=32\pi\ell_p^2$  (grey dashed lines are at the intersection with the usually allowed (positive half-integers) values of the spin $j_{min}=1/2,1,3/2,\dots$). Furthermore, we set the energy scale to $\Delta p\simeq m_p$. For $A=16\pi\ell_p^2$, constraints on non-extensivity discussed above turn into the condition $|\beta|\sim\mathcal{O}(10^{-2})$, which is  consistent with the result in Sec~\ref{Verlinde} (see below Eq.~\eqref{betaest2}). Notice that this estimate is slightly reduced in the case of  $A=32\pi\ell_p^2$. In fact, the $\beta$ range can be shortened by increasing the area of the black hole sufficiently. This implies that, for any physically realizable black hole, a small departure from extensivity/HUP-based QM  
is likely to account for a minimum spin $j_{min}\neq1$. However, it must be observed that only the $\beta<0$ scenario can accommodate 
$j_{min}=1/2$, since $j_{min}>1$ as far as one considers positive values of $\beta$ (see Fig.~\ref{miva}). Therefore, contrary to the result of~\cite{QNM}, we conclude that both SU(2) and SO(3) are possible gauge groups of the spin networks of LQG in the presence of GUP.

\subsubsection{Immirzi parameter}
\begin{figure}[t]
\begin{center}
\includegraphics[width=8.4cm]{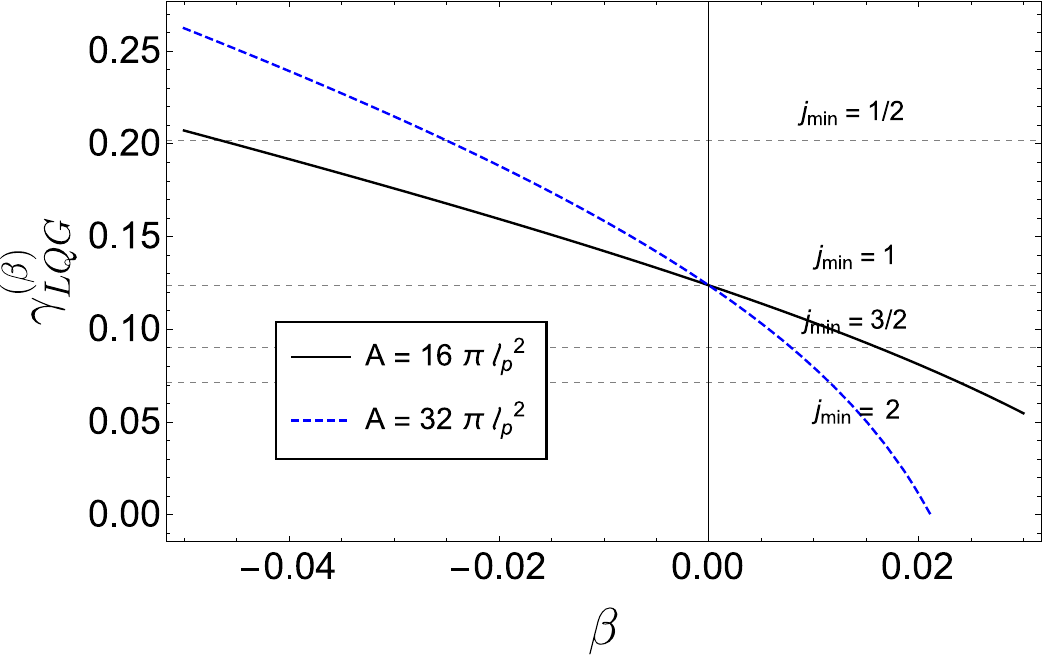}
\caption{Plot of $\gamma^{(\beta)}_{LQG}$ as a function of $\beta$ for $A=16\pi\ell_p^2$ (black solid line) and $A=32\pi\ell_p^2$ (blue dashed line). We set the energy scale to $\Delta p\simeq m_p$ (online colors).}
\label{GUPTS2}
\end{center}
\end{figure}

Based on Eqs.~\eqref{gLQG} and~\eqref{jminb}, 
let us also exhibit the GUP-corrected expression of Immirzi parameter
\begin{eqnarray}
    \gamma^{(\beta)}_{LQG}&=&\frac{\log 3}{\pi} \times \nonumber\\[2mm]
    &&\hspace{-1.5cm}\left\{\left[1-\beta\,\frac{A \Delta p^2\, m_p^2}{2\hslash^2\left(m_p^2+3\beta \Delta p^2\right)}\right]^{-\frac{4\hslash^2\log 3\,\left(m_p^2+3\beta\Delta p^2\right)}{\beta\,A\Delta p^2\,m_p^2}}\!\!-1\right\}^{-\frac{1}{2}}\!\!\!\!\!.  \nonumber\\[2mm]
\end{eqnarray}
The behavior of $\gamma^{(\beta)}_{LQG}$ versus $\beta$ is plotted in Fig.~\ref{GUPTS2} for $A=16\pi\ell_p^2$ and $A=32\pi\ell_p^2$.
Some comments are in order here. Though seemingly counter-intuitive, we find that the modified Immirzi parameter somehow depends on the properties of the state through $\Delta p$.
A possible explanation
is that since in our framework Tsallis statistics affects quantum mechanics in a state-dependent way (see Eq.~\eqref{17.ccd}), it is then quite natural to expect that the counting of micro-states for a given system is such. In turn, given that the Immirzi parameter is fixed by matching the semiclassical black hole entropy and the counting of micro-states in LQG, one gets the conclusion. Furthermore, we notice that $\gamma^{(\beta)}_{LQG}$ correctly equals the expected limit~\eqref{standgamma} for $\beta\rightarrow0$. Finally, 
the value $\gamma^{(\beta)}_{LQG}=\frac{\log 3}{\pi\sqrt{3}}$, corresponding to
$j_{min}=1/2$, is obtained for small (negative)
deviation of $\beta$ from zero for any realizable black hole, consistently with what is stated above.

\subsection{Magueijo--Smolin DSR \label{DSR}}
\label{DSRI}

It is interesting to note that the deformed commutation relation (\ref{comm}) with $\beta <0$ and magnitude~(\ref{betaest2})
complies nicely with Magueijo--Smolin DSR~\cite{MS1,MS2}.
In a nutshell, DSR is a theory that coherently tries to implement a second invariant (namely $m_p$ or equivalently $\ell_p$)
besides the speed of light into the transformations among inertial reference frames. The Magueijo--Smolin DSR model predicts
that the DCR should vanish at Planck scale (thus physics should become deterministic there), while at low energies it approaches the
conventional canonical commutator. In our case,
we indeed see from (\ref{betaest2}) that, for allowed cosmological times, the deformation parameter $|\beta(t)|$ monotonically decreases with increasing $t$.
This is consistent with the expectation of the Magueijo--Smolin DSR model
that $\beta$ should grow at high-energy scales
so that at Planck scale the right-hand-side of the commutation relation~\eqref{comm}
vanishes. Although such a behavior is quite encouraging, it is only indicative, as we cannot extend the validity of our formalism up to the Planck scale, since this would be at odds with the presumed quasi-classical domain of validity.

Nevertheless, a quantitative comparison can still be done by employing the DCR predicted by the Magueijo--Smolin model. To keep our reasoning simple, it suffices to focus on the one-dimensional problem, since in higher spatial dimensions the DSR-inspired commutation relations are compatible with spatial commutativity. To make the generalization introduced in Sec.~\ref{3D generalization} consistent with the above scenario, one should, in principle, account for an extra term in Eq.~\eqref{79.aa} which still preserves rotational invariance and complies with a precise prescription (as discussed, e.g., in Ref.~\cite{angmom2}). However, for the purpose of the upcoming considerations, such a technical aspect can be safely neglected, as it would not affect the validity of our result. 

Let us introduce the Magueijo--Smolin deformed commutation relation~\cite{MS2}
\begin{equation}\label{magsmo}
\left[\hat{x},\hat{p}\right]\ = \ i\left(1-\ell_p \hat{E}\right)\, ,   
\end{equation}
with $\hat{E}$ denoting the energy operator. A straightforward comparison with Eq.~\eqref{comm} and $\beta<0$ shows that the two DCRs can be related. Indeed, by reconstructing the energy of a non-relativistic system in Eq.~\eqref{comm} by suitably introducing the mass of the analyzed quantum system, we immediately deduce that the value of the deformation parameter providing the exact match of the prediction associated to the two distinct pictures is given by
\begin{equation}\label{betam}
\beta\equiv\beta(m) \ = \ \frac{m_p}{2m}\, .
\end{equation}
One can also achieve this result by examining the non-relativistic limit of the DSR-inspired commutation relations~\cite{coraddu}, and then comparing it to the GUP case.

Eq.~(\ref{betam}) encodes an interesting outcome. As a matter of fact, in the framework of DSR, the invariance of the scale (i.e., Planck scale) at which quantum gravitational effects are deemed to become relevant requires a fundamental deformation of the action of Lorentz transformations on momentum space. In particular, the ensuing modified composition laws acquire a non-linear character~\cite{MS1,MS2}, which renders the whole analysis highly non-trivial. A consequence of this occurrence that is commonly encountered in several quantum gravity candidate models is related to the possibility of witnessing a non-vanishing curvature in momentum space (see for instance Refs.~\cite{curved1,curved2,curved3,curved4,curved5,curved6,curved7}), which is thus liable to be treated with geometric tools stemming, e.g., from general relativistic methods. In light of this observation, a first resemblance between the Magueijo--Smolin DSR and GUP can already be drawn at this stage. Indeed, in a series of recent works \cite{fab1,fab2} it was shown that GUP-inspired deformations of the canonical Heisenberg algebra can be reinterpreted in terms of a non-vanishing curvature in momentum space, thereby making the aforesaid approach compatible with DSR also from this standpoint.

On the other hand, the most compelling subject paired with Eq.~\eqref{betam} and the DSR theoretical scheme is represented by the so-called ``soccer ball'' problem~\cite{sb1,sb2,sb3,sb4,sb5,sb6}. In short, the issue lies in the non-linear behavior of momentum composition law, which in principle does not prohibit the enhancement of quantum gravitational corrections when composite systems are accounted for; the more macroscopic the considered system, the more pronounced this effect is expected to become. However, since the macroscopic world we experience everyday exhibits no trace of quantum gravitational signatures, there should be an explanation that motivates the suppression of such a phenomenon. 
In this sense, Eq.~\eqref{betam} might be viewed as a potential way out, since the inverse proportionality between $\beta$ and $m$ conveys a reduction in the magnitude of the deformations to the standard Heisenberg algebra when the mass of the studied system increases (i.e., when the mesoscopic/macroscopic regime is approached). This ``inverse soccer ball'' tendency is not entirely new in the context of GUP, as it has already been pointed out 
in Ref.~\cite{camel}. Therefore, the problem originated by the non-linear momentum composition law might be solved by requiring consistency between the predictions of the two distinct deformations of the standard Heisenberg algebra. In so doing, we would manage to achieve a foreseeable scaling dependence on the size of the considered system regardless of the non-triviality of the underlying momentum space, which would thus be left untouched. More details on this topic require further investigation and will be presented elsewhere.

\section{Discussion and Conclusions}
\label{Conc}

In this paper, we have unified two seemingly unrelated concepts, namely the generalized uncertainty principle and Tsallis (thermo)statistics. On the one hand, the GUP strives to explore the consequences of the existence of a minimal length scale (be it fundamental or emergent). For this reason, it has a large number of both theoretical and experimental implications, which range from early Universe cosmology and astrophysics, to condensed matter theory and quantum optics. On the other hand, Tsallis statistics is a theoretical concept that accounts for systems with long-range correlations or long-time memory, for which the conventional central limit theorem does not apply.  Again, it has a large number of both theoretical and experimental implications that span such fields as statistical physics, thermodynamics and complexity theory, with applications ranging from condensed matter theory, fluid dynamics and social sciences to quantum information. A merger of the two concepts presented here is intriguing from a conceptual perspective for the following reasons: 
\begin{itemize}
  \item Modifications of quantum mechanics caused by the existence of a minimal length scale (be it fundamental or effective) is in the semi-classical (decoherence) limit necessarily equivalent to Tsallis statistics. This, in turn, provides a new methodology for the study of GUP systems and, at the same time, a new arena for Tsallis statistics. 
  \item Quantum mechanical systems in the decoherence limit represent an ongoing field of intense research in quantum information theory because decoherence is the main impediment to the realization of devices for quantum information processing, e.g., quantum computers. It is thus important to get a handle on how prospective GUP modifications to quantum mechanics would influence decoherence-borne errors in quantum information processing and sensing.
  \item Decohered quantum regime is also known to be pertinent in observational cosmology  (decoherence description is supposed to be valid at the late-inflation period --- after the first Hubble radius crossing, and perhaps even after its end during reheating). Prospective GUP corrections should be then relevant at these early epochs of the Universe.
    \item In Sec.~\ref{relat} we have noticed that the form~\eqref{1.cf} of the GUP 
    provides the leading-order approximation towards Maxwell--J\"{u}ttner (i.e. relativistic) type deformation of the uncertainty relation. However, it must be said that Maxwell--J\"{u}ttner distribution only represents the first naive approach to develop relativistic statistical mechanics, as it arises from  Maxwell--Boltzmann distribution with the classical energy-velocity relation being replaced by its relativistic counterpart. On the other hand, a self-consistent relativistic statistical theory is one built upon the Kaniadakis probability density function and ensuing entropy~\cite{Kaniadakis1,Kaniadakis2,Kaniadakis3}. The latter represents a one-parameter modification of  Boltzmann--Gibbs--Shannon entropy and it is naturally imposed by Lorentz transformations (see also~\cite{RevLucia} for a recent review of gravitational and cosmological applications in Kaniadakis statistics).  Based on our result, Kaniadakis' probability amplitude should then coincide with coherent states associated with a fully relativistic version of GUP. Since GUP models in relativistic theories are still controversial, we expect that 
  the recipe presented here could pave the way for the correct formulation of relativistic GUP
  when applied to the Kaniadakis statistics.
\end{itemize}
At the same time, advances prompted by the present analysis are also expected in more practical/experimental contexts, since:
\begin{itemize}
  \item Tsallis entropy with its entropy power are measurable quantities (there is a number of coding theorems and communication protocols for them) and they are indeed routinely used both in classical and quantum information theory and in quantum optics. So, the Tsallis entropy-power uncertainty relations discussed here are experimentally accessible.
  \item Inflation-based considerations discussed here not only restrict a numerical value of the GUP deformation parameter
  $\beta$, but the GUP semi-classical regime predicts, e.g., a very specific polytrope state equation for inflaton field.  A specific imprint of this should be observed in the cosmic microwave background radiation~\cite{Bar:16,Bar:18}.
  \item Postulating a similar GUP commutator between the canonical variables of the electromagnetic field in quantum optics, one can evaluate corrections to the radiation pressure noise and shot noise in various optomechanical systems in their semi-classical regime, e.g., Michelson--Morley type interferometers. These corrections might be experimentally observed, e.g., in future advanced LIGO detectors.
\end{itemize}

To substantiate our point, we have employed here the NTT to generalize  Verlinde's entropic force. Apart from obtaining a modified Newtonian (basically Mannheim--Kazanas) potential, we have argued that 
such a generalization should be phenomenologically pertinent at the late-inflation epoch. The corresponding dependence of the GUP $\beta$ parameter on cosmological time $t$ has also been derived for the reheating epoch. The $\beta$ parameter inferred in this way is consistent both with values predicted by string-theory models and with the naturalness principle. Moreover,
we have shown that the dependence of $\beta$ on $t$ is compatible with the  Magueijo--Smolin doubly special relativity scenario. Moreover, a more precise comparison has revealed more similarities between the two  approaches, such as the common prediction of an underlying curved momentum space. Interestingly, these contact points might potentially represent the solution to a phenomenological issue called ``soccer ball'' problem plaguing DSR, without sacrificing the non-trivial geometry of momentum space.
Finally, within the context of the NTT, we have derived new $\beta$-dependent expressions for the lowest possible value of the spin and Immirzi parameter in Loop Quantum Gravity. We have shown that the $\beta<0$ choice can easily accommodate $j_{min}=1/2$, while $j_{min}>1$ provided one works with positive values of $\beta$. This implies that both SU(2) (which has been normally adopted) and SO(3) (which has been claimed in~\cite{QNM}) are possible gauge groups of LQG spin networks in the presence of GUP.

\medskip



\begin{acknowledgments}
We thank John Klauder,  Vladimir Lotoreichik
and Pavel Exner for useful discussions and comments on the manuscript. 
P.J. was in part supported by the FNSPE CTU grant RVO14000. G.G.L. acknowledges the Spanish ``Ministerio de Universidades'' for the awarded Maria Zambrano fellowship and funding received
from the European Union - NextGenerationEU. L.P. is grateful to the ``Angelo Della Riccia'' foundation for the awarded fellowship received to support the study at Universit\"at Ulm. G.G.L. and L.P. acknowledge networking support by the COST Action CA18108.
\end{acknowledgments}

\appendix

\section{Eigenstates of the position operator \label{Appendix A}}

Since the generalized commutator between position and momentum takes the form \eqref{comm}, the plane waves are no longer eigenstates of the operator $\hat{x}$ in the momentum representation and of $\hat{p}$ in the position representation. This in turn means  that wave functions in position and momentum representation are not connected via a Fourier transform. To see how they are related, let us first consider the eigenstates
of the operator $\hat{x}$ in the momentum representation. These are given by solving the following eigenvalue equation
\begin{equation}
\hat{x}\langle p|x\rangle \, \equiv \, x \psi_x(p) \, = \, i\hbar \left(\frac{d}{dp} + \frac{\beta}{m^2_p}\, p^2 \frac{d}{dp} + \frac{\beta}{m^2_p} p\right)\psi_x(p)\, ,
\end{equation}
which can be equivalently rewritten as
\begin{eqnarray}
\frac{d}{dp} \ \! \psi_x(p) \, = \, \frac{\left(x \, - \, i\hbar \frac{\beta}{m^2_p} p\right)}{i\hbar\left(1 \, + \, \frac{\beta}{m^2_p} p^2 \right)} \ \! \psi_x(p) \, .
\end{eqnarray}
The solution is of the form
\begin{eqnarray}
\psi_x(p) \, = \, A_x \ \! \frac{e^{-ixm_p\arctan\left(p \sqrt{\beta}/m_p  \right)/\hbar\sqrt{\beta}}}{\sqrt{m^2_p  \, + \, p^2 \beta}}\, ,
\label{3aa}
\end{eqnarray}
for positive $\beta$, and
\begin{eqnarray}
\psi_x(p) \, = \, B_x \ \!
\frac{e^{-ixm_p{\mbox{\scriptsize{arctanh}}}\left(p \sqrt{|\beta|}/m_p  \right)/\hbar\sqrt{|\beta|}}}{\sqrt{m^2_p  \, - \, p^2 |\beta|}}\, ,
\label{3ab}
\end{eqnarray}
for negative $\beta$ (i.e., $\beta = -|\beta|$).

In the following, we discuss the two cases $\beta >0$ and $\beta <0$ separately.

\subsection*{Negative $\beta$ case}
\label{negbetaEigen}

In this case, $\psi_x(p)$  is not quadratically integrable, since 
\begin{eqnarray}
\nonumber
|\!| \psi_x|\!|^2 &=& \langle x|  x\rangle \, =  \, |B_x|^2 \int_{-m_p/\sqrt{|\beta|}}^{m_p/\sqrt{|\beta|}} \frac{dp }{m_p^2 - p^2|\beta|}\\[2mm]
\nonumber
& =&
\{ z =  m_p\ \! \mbox{{arctanh}}\left(p \sqrt{|\beta|}/m_p  \right)/\sqrt{|\beta|}\}\nonumber \\[2mm]
& =& \left(\frac{|B_x|}{m_p}\right)^2 \int_{-\infty}^{\infty} dz \ = \ \infty\, .
\end{eqnarray}
The ensuing scalar product for two eigenstates is
\begin{eqnarray}
\nonumber
&&\hspace{-0.4cm}\langle x'|  x\rangle\\[2mm] 
\nonumber
&&\hspace{-0.4cm}={|B_x|}^2 \int_{-m_p/\sqrt{|\beta|}}^{m_p/\sqrt{|\beta|}}dp \ \! \frac{ e^{-i(x-x')m_p{\mbox{\scriptsize{arctanh}}}\left(p \sqrt{|\beta|}/m_p  \right)/\hbar\sqrt{|\beta|}}}{m_p^2 - p^2|\beta|} \\[2mm]
\nonumber
&&\hspace{-0.4cm}= \left(\frac{|B_x|}{m_p}\right)^2 \int_{-\infty}^{\infty} dz \ e^{-i(x-x')z/\hbar}\nonumber \\[2mm] &&\hspace{-0.4cm}= \left(\frac{|B_x|}{m_p}\right)^2 2\pi \hbar \ \!\delta(x-x')\, .
\end{eqnarray}
Therefore, we can set $B_x = \sqrt{m_p^2/2\pi \hbar}$. 
In Ref~\cite{JLLP} it was shown that the $\hat{x}$ operator is self-adjoint when an appropriate (and in a sense natural) domain is chosen.

In passing, we should note that, although continuous observables such
as $\hat{x}$ or $\hat{p}$ are routinely employed in quantum theory, they are
really unphysical idealizations: the set of possible
outcomes in any realistic measurement is always
countable, since the state space of any apparatus
with finite spatial extent has a countable
basis.
%
%
In turn, our reasoning related
to $\beta < 0$ should thus be understood in this mathematically idealized sense --- as done with conventional Heisenberg $p$-$x$ uncertainty relations.

\section{Connection between wave functions in momentum and position representation}
\label{ConWavF}

In this Appendix, we discuss the connection between wave functions in momentum and position representation in the presence of the modified commutator~\eqref{comm}.

\subsection*{Negative $\beta$ case}
\label{negbetaConn}

In this case the position and momentum representations of a wave function are related via the relation
\begin{eqnarray}
\psi(x) &\hspace{-1mm}=\hspace{-1mm}&  \int_{\frac{-m_p}{\sqrt{|\beta|}}}^{\frac{m_p}{\sqrt{|\beta|}}} \frac{dp}{\sqrt{2\pi \hbar}} \ \! \frac{e^{ixm_p{\mbox{\scriptsize{arctanh}}}\left(p \sqrt{|\beta|}/m_p  \right)/\hbar\sqrt{|\beta|}}}{\sqrt{1   -  p^2 |\beta|/m_p^2}}\ \! \tilde{\psi}(p)\nonumber \\[2mm]
&\hspace{-1mm}=\hspace{-1mm}&  \{ z =  m_p\mbox{{arctanh}}\left(p \sqrt{|\beta|}/m_p \right)/\sqrt{|\beta|}\}\nonumber \\[2mm]
&\hspace{-1mm}=\hspace{-1mm}&  \int_{\mathbb{R}} \frac{dz}{\sqrt{2\pi \hbar}} \ \! e^{ixz/\hbar} \ \! \frac{\tilde{\psi}(m_p\tanh(z \sqrt{|\beta|}/m_p)/\sqrt{|\beta|})}{\cosh(z \sqrt{|\beta|}/m_p)} \nonumber
 \\[2mm]
&\hspace{-1mm}=\hspace{-1mm}&  \int_{\mathbb{R}} \frac{dz}{\sqrt{2\pi \hbar}} \ \! e^{ixz/\hbar} \bar{\psi}(z)   \, ,
\label{B.7cd}
\end{eqnarray}
where 
\begin{equation}
   \bar{\psi}(z) \ = \  \frac{\tilde{\psi}(m_p\tanh(z \sqrt{|\beta|}/m_p)/\sqrt{|\beta|})}{ {\cosh(z \sqrt{|\beta|}/m_p)} }.
\end{equation}
Note that this formula is valid only for $D=1$ dimension. In passing we can easily check that the analogue of  Parseval--Plancherel theorem holds, namely
\begin{eqnarray}
\nonumber
\hspace{-4mm}\int_{\mathbb{R}} dx \ \! |\psi(x)|^2 &=&  \int_{-m_p/\sqrt{|\beta|}}^{m_p/\sqrt{|\beta|}} {d p} \ \! |\tilde{\psi}(p)|^2 \nonumber  \\[2mm] &=& \ \int_{\mathbb{R}} {d z} \ \! |\bar{\psi}(z)|^2  \, ,
\end{eqnarray}
or equivalently $|\!| \psi|\!|_2 \ = \ |\!| \tilde{\psi}|\!|_2 \ = \ |\!| \bar{\psi}|\!|_2 $.

Note also that from the last line in (\ref{B.7cd}), one can also easily deduce that the momentum operator in the position representation has the form
\begin{eqnarray}
\hat{p}^{(x)} \ = \ m_p \tanh\left(-i \hbar \sqrt{|\beta|}/m_p \ \! \frac{d}{dx} \right)/\sqrt{|\beta|}\, .
\end{eqnarray}
It is straightforward to verify that the operator satisfies the canonical commutation relation~\eqref{comm}.

There is yet another interesting consequence of Eq.~(\ref{B.7cd}), namely one can directly compute from it the corresponding {\em position-space coherent state}. In particular, by using the Tsallis probability amplitude {(\ref{sol})} (i.e., momentum-space coherent state) we can write for the corresponding position-space coherent state $\psi_{CS}(x)$ that
\begin{eqnarray}
\nonumber
&&\hspace{-6mm}\psi_{CS}(x) \nonumber \\[2mm]
&&\hspace{-3mm}= \ N \int_{\mathbb{R}} \frac{dz}{\sqrt{2\pi \hbar}} \ \! e^{ixz/\hbar} \nonumber \\[2mm]
&&\times\,\frac{\left[m^2_p - m^2_p (\tanh(z \sqrt{|\beta|}/m_p))^2 \right]^{m_p^2/(2|\beta| \gamma \hbar) - 1/2}}{\cosh\left(z \sqrt{|\beta|}/m_p \right)}\nonumber \\[2mm]
&&\hspace{-3mm}= \ N m_p^{m_p^2/(|\beta| \gamma \hbar)-1} \int_{\mathbb{R}} \frac{dz}{\sqrt{2\pi \hbar}} \ \! e^{ixz/\hbar} \nonumber \\[2mm] 
&&\times\,{\cosh\left(z \sqrt{|\beta|}/m_p \right)^{-m_p^2/(|\beta| \gamma \hbar)}}\nonumber \\[2mm]
&&\hspace{-3mm}= \ \tilde{N }
\left|\Gamma\left(\frac{m^2_p/(|\beta| \gamma \hbar)}{2} + i \frac{xm_p}{2\sqrt{|\beta|}\hbar}\right)\right|^2  \nonumber \\[2mm]
&&\hspace{-3mm} = \ \sqrt{\frac{m_p \ \!\Gamma(2 m^2_p/(|\beta| \gamma \hbar))}{4\pi \sqrt{|\beta|} \hbar  \ \! \Gamma^4(m^2_p/(|\beta| \gamma \hbar))}}\nonumber \\[2mm]
&&\times\,\left|\Gamma\left(\frac{m^2_p/(|\beta| \gamma \hbar)}{2} + i \frac{xm_p}{2\sqrt{|\beta|}\hbar}\right)\right|^2 \, .
\label{CS-x}
\end{eqnarray}
In the derivation we used 
the Ramanujan formula~\cite{Ramanujan}
\begin{eqnarray}
&&\mbox{\hspace{-15mm}}\int_{-\infty}^{\infty} e^{-i\xi s} \ \! \left|\Gamma(a + is) \right|^2 \ \! ds \nonumber \\[2mm]
&&\mbox{\hspace{3mm}}= \ 
\frac{2\pi}{2^{2a}}  \Gamma(2a) \left[ \cosh(\xi/2)\right]^{-2a}  \, ,
\end{eqnarray}
that is valid for $a\in (-1,0)\cup (0,\infty)$. The normalization factor on the last line of (\ref{CS-x}) was obtained by employing the Mellin--Barnes beta integral (cf. e.g. Refs.~\cite{distributions,Ober})
\begin{eqnarray}
\int_{-\infty}^{\infty} \ \! \left|\Gamma(a + ib s) \right|^4 \ \! ds \ = \ \frac{2 \pi}{b} \frac{\Gamma^4(2a)}{\Gamma(4a)}\, .
\label{19.SM.aa}
\end{eqnarray}
In passing we note that the state $\psi_{CS}(x)$ is even-parity state (as required) and, in addition,  it belongs to the Schwartz class, i.e., it decays rapidly at infinity along with all derivatives.

For consistency we can now check that $\psi_{CS}(x)$ from Eq.~(\ref{CS-x}) provides a correct positional variance, which together with the momentum variance deduced from  $\tilde{\psi}_{CS}(p)$ (cf. {Eq.~(\ref{19.bbb})}) saturates the GUP~\eqref{gup}.
To this end one can use the formula for the Fourier transform of  $\left|\Gamma(a + ib s) \right|^4$ (see~\cite{Ober}, Eq.~(274), p. 46) to show that
\begin{eqnarray}
\mbox{\hspace{-4mm}}\frac{b \Gamma(4a)}{2\pi \Gamma^4(2a)}\int_{-\infty}^{\infty} \ \! s^2 \ \! \left|\Gamma(a + ib s) \right|^4 \ \! ds =  \frac{a^2}{b^2(1+4a)}\, .
\end{eqnarray}
If we now use from (\ref{19.SM.aa}) that $a = m_p^2/(2|\beta|\gamma \hbar)$ and $b = m_p/(2\sqrt{|\beta|}\hbar)$ we get
\begin{eqnarray}
\nonumber
(\Delta x)^2_{CS}  &=&  \int_{-\infty}^{\infty} \ \! x^2 \ \! \psi_{CS}(x) \ \! dx  \\[2mm]
&=&  \frac{\hbar m_p^2}{2m_p^2 \gamma + \hbar |\beta| \gamma^2} \ = \ \frac{(\Delta p)_{CS}^2}{\gamma^2}\, ,
\label{SM.40.cc}
 \end{eqnarray}
where the last identity results from (\ref{18cd}) and {(\ref{var2})}. This is equivalent to the saturated GUP.

\section{Derivation of the measure $\mu(x_0, p_0)$ \label{Appendix C}}

Here we derive the  measure for the resolution of unity (\ref{c.11.cfftext}). 
To this end we realize that the expression
\begin{eqnarray}
F(a/2,2z) \ \equiv \ \int_{-a/2}^{a/2} dy \ \! {\tilde{\mu}(y)} \ \! |\tilde{N}(y)|^2 \ \!e^{2yz}\, ,
\label{C1.bb.f}
\end{eqnarray}
is the {\em finite Laplace transform}. 
The inversion formula for the finite Laplace transform is  given by the principal value integral~\cite{Laplace-transf}
\begin{eqnarray}
&&\mbox{\hspace{-10mm}}{\tilde{\mu}(y)} \ \! |\tilde{N}(y)|^2 \nonumber \\[2mm] &&= \ \frac{1}{2\pi i} \lim_{R \rightarrow \infty} \int_{c - iR}^{c + iR} e^{-sy}\ \! F(a/2,s) \ \! ds\, .
\label{C2.bb}
\end{eqnarray}
Here the integral is taken over any open contour $\Gamma$ joining the points
$c - iR$ and $c + iR$ in the finite complex $s$ plane as $R \rightarrow \infty$. 
The arbitrariness of $\Gamma$ stems from the fact that $F(a/2,s)$ is an entire function of $s$.
In practice, the integration is typically done along the vertical line $Re(s) = c$ in the complex plane. In contrast to the usual inverse Laplace transform, the value of $c$ can be freely chosen. So, in particular, (\ref{C2.bb}) is not necessarily related to the Bromwich integral. 
%
%
%

If we now demand that
\begin{eqnarray}
F(a/2,2z) &=& \frac{m_p}{2\pi \hbar^2 \gamma \sqrt{|\beta|}} \ \! (\cosh z)^a 
\, ,
\label{C3.bb}
\end{eqnarray}
we should obtain, by inserting (\ref{C3.bb}) to (\ref{C2.bb}), the desired measure $\mu(x_0,p_0)$. The actual integral that needs to be evaluated in order to obtain the measure is 
\begin{eqnarray}
\frac{1}{2\pi i} \lim_{R \rightarrow \infty} \int_{c - iR}^{c + iR} e^{-sy}\ \! \left(\cosh \mbox{$\frac{s}{2}$}\right)^a \ \! ds\, .
\label{C4.bb}
\end{eqnarray}
By choosing $c=0$, we can rewrite (\ref{C4.bb}) in the form
\begin{eqnarray}
&&\mbox{\hspace{-4.5mm}}
\frac{1}{\pi} \lim_{R \rightarrow \infty} \int_{-R}^{R} e^{-i2sy}\ \! \left(\cos {s}\right)^a \ \! ds\nonumber \\[2mm]
&&\mbox{\hspace{-4.5mm}}=  \frac{1}{\pi} \lim_{n \rightarrow \infty} \sum_{k= -n}^n \int_{(k-1/2)\pi}^{(k+1/2)\pi} e^{-i2sy}\ \! \left(\cos s\right)^a \ \! ds\nonumber \\[2mm]
&&\mbox{\hspace{-4.5mm}}=  \frac{1}{\pi} \lim_{n \rightarrow \infty} \sum_{k= -n}^n (-1)^{ka} e^{-i2k\pi y} \int_{-\pi/2}^{\pi/2} e^{-i2sy}\ \! \left(\cos s\right)^a \ \! ds\nonumber \\[2mm]
&&\mbox{\hspace{-4.5mm}}= \frac{1}{2^a} \lim_{n \rightarrow \infty} \sum_{k= -n}^n e^{-ik\pi (2y \pm a)} \frac{\Gamma(a+1)}{\Gamma(1+ \mbox{$\frac{a}{2}$} + y)\Gamma(1+ \mbox{$\frac{a}{2}$} - y)}\nonumber \\[2mm]
&&\mbox{\hspace{-4.5mm}}=  \frac{1}{2^{a-1}}  \!\sum_{k= -\infty}^{\infty} \delta(\mbox{$\pm 2y +  a - 2k$}) \frac{\Gamma(a+1)}{\Gamma(1+ \mbox{$\frac{a}{2}$} + y)\Gamma(1+ \mbox{$\frac{a}{2}$} - y)}\nonumber \\[2mm]
&&\mbox{\hspace{-4.5mm}}=  \frac{1}{2^{a-1}}  \sum_{k= 0}^{\infty}\ \!\delta(\mbox{$\pm 2y + a - 2k$}) \frac{\Gamma(a+1)}{\Gamma(1 + k)\Gamma(1+ a - k )},~~~~
\label{C5.bb.c}
\end{eqnarray}
where on the second line we have set $R = (n+1/2)\pi$, on the fourth line we have used the identity~\cite{ramanujan_b}
\begin{eqnarray}
&&\mbox{\hspace{-15mm}}
\int_{-\pi/2}^{\pi/2} e^{-isy}\ \! \left(\cos s\right)^a \ \! ds \nonumber \\[2mm] 
&&= \ \frac{\pi}{2^a} \frac{\Gamma(a+1)}{\Gamma[1 + \mbox{$\frac{1}{2}$}(a+y)]\Gamma[1 + \mbox{$\frac{1}{2}$}(a-y)]}\nonumber \\[2mm] 
&& = \  \frac{\pi}{2^a} \binom{a}{\mbox{$\frac{1}{2}$}(a+y)}
\, ,
\end{eqnarray}
(the last line represents the generalised binomial coefficient),
on the fifth line we have employed Poisson's summation formula~\cite{vladimirov}
\begin{eqnarray}
\sum_{k=-\infty}^{\infty} \delta(x-kT) \ = \ \frac{1}{T} \sum_{k=-\infty}^{\infty} \exp\left(\pm i \frac{2\pi}{T} k x\right)\, ,
\end{eqnarray}
and finally, on the last line, we have taken advantage of the fact that $\Gamma(1+k)$ has poles when $k$ is a negative integer. 
%
%

By inserting (\ref{C5.bb.c}) into  (\ref{C1.bb.f}), we can verify that the desired $(\cosh z)^a$ term is indeed the solution of the integral. To see this explicitly, consider the upper sign in (\ref{C5.bb.c}). We first notice that  (\ref{C5.bb.c}) can be rewritten as
\begin{eqnarray}
\frac{1}{2^a} \sum_{k=0}^{\infty} e^{(a-2k) \mbox{$\frac{\partial}{\partial 2y}$}} \binom{a}{k} \ \! \delta(y)  =  \left[\cosh\left(\mbox{$\frac{1}{2}\frac{\partial}{\partial y}$}\right)\right]^a  \delta(y)\, .~~~
\label{c8.vb}
\end{eqnarray}
By coupling this result with the identity~\cite{vladimirov}
\begin{eqnarray}
e^{2zy} \left[\cosh\left(\mbox{$\frac{1}{2}\frac{\partial}{\partial y}$}\right)\right]^a \ \! \delta(y) &=& 
\delta(y) \ \! e^{zy} \ \! (\cosh z)^a \nonumber \\[2mm]
&=&  
\delta(y)  \ \! (\cosh z)^a\, ,
\end{eqnarray}
[which can be checked term-by-term by Taylor expanding the exponent in (\ref{c8.vb})], we can write (\ref{C1.bb.f}) as 
\begin{eqnarray}
&&\mbox{\hspace{-12mm}}\int_{-a/2}^{a/2} dy \ \! {\tilde{\mu}(y)} \ \! |\tilde{N}(y)|^2 \ \!e^{2yz} \nonumber \\[2mm]
&&\mbox{\hspace{-4mm}} = \
\frac{m_p}{i (2\pi)^2 \hbar^2 \gamma \sqrt{|\beta|}} \int_{-a/2}^{a/2} \!dy \ \! e^{2yz} \nonumber \\[2mm] 
&&\mbox{\hspace{1mm}}\times \ \lim_{R \rightarrow \infty} \int_{c - iR}^{c + iR} \!ds \ \!e^{-sy}\ \! \left(\cosh \mbox{$\frac{s}{2}$}\right)^a  \nonumber \\[2mm]
&&\mbox{\hspace{-4mm}}  = \ \frac{m_p}{ 2\pi \hbar^2 \gamma \sqrt{|\beta|}} \int_{-a/2}^{a/2} dy \ \! \delta(y)  \ \! (\cosh z)^a \nonumber \\[2mm]
&&\mbox{\hspace{-4mm}}  = \ \frac{m_p}{ 2\pi \hbar^2 \gamma \sqrt{|\beta|}}  \ \! (\cosh z)^a 
\, ,
\label{C10.bb.f}
\end{eqnarray}
which is indeed $F(a/2,2z)$. We would get the same result if we had begun with the lower sign in (\ref{C5.bb.c}).

In view of the explicit form of $|\tilde{N}(y)|^2$, see Eq.~(\ref{37.ccf}), we get for the measure  
\begin{eqnarray}
&&\mbox{\hspace{-10mm}}\mu(x_0,p_0) \ = \ \mu(p_0)  \ = \ \tilde{\mu}(y) \nonumber \\[2mm]
&&\mbox{\hspace{-5mm}}= \ \frac{1}{2\pi\hbar} \frac{1}{\left[1-\left(\frac{2y}{a}\right)^{\!2}\right]} \ \! \sum_{k=0}^{\infty} \delta(\pm y + a/2 -k)\, ,
\label{c.11.cff}
\end{eqnarray}
with $y \in (-a/2,a/2)$.
In (\ref{c.11.cff}), any sign convention can be used. 
The resolution of unity thus reads
\begin{eqnarray}
&&\mbox{\hspace{-8.5mm}}\mathds{1}  \ = \ \int_{\mathds{R}} dx_0 \int_{-m_p/\sqrt{|\beta|}}^{m_p/\sqrt{|\beta|}} dp_0 \ \! \mu(p_0) \ \! |\psi, p_0,x_0\rangle\langle \psi, x_0, p_0|\nonumber \\[2mm]
&&\mbox{\hspace{-5mm}}= \ \frac{\hbar \gamma\sqrt{|\beta|}}{m_p}  \int_{\mathds{R}} dx_0 \int_{-a/2}^{a/2} dy \ \!\tilde{\mu}(y) \ \! |\psi, y,x_0\rangle\langle \psi, x_0, y|\nonumber \\[2mm]
&&\mbox{\hspace{-5mm}}= \ \int_{\mathds{R}} dx_0 \sum_{k=0} \ \!\frac{\hbar \gamma\sqrt{|\beta|}}{2\pi \hbar \ \!m_p}  \ \!\frac{\left(\frac{a}{2}\right)^2}{k(a-k)}\ \! |\psi, k,x_0\rangle\langle \psi, x_0, k|\nonumber \\[2mm]
&&\mbox{\hspace{-5mm}}= \ \int_{\mathds{R}} dx_0 \sum_{p_0} \ \! \frac{\Delta p_0}{2\pi \hbar} \  \frac{|\psi, p_0,x_0\rangle\langle \psi, x_0, p_0|}{\left[1- \frac{p_0^2 |\beta|}{m_p^2}   \right]} \, .
\label{c.12.cc}
\end{eqnarray}
Since $\delta(\pm y + a/2 -k)$ implies that only contributing values of $p_0$ are $\mp m_p/\sqrt{|\beta|} \pm k \gamma \hbar \sqrt{|\beta|}/m_p $, we have that the difference between two neighbouring values of $p_0$ is 
\begin{eqnarray}
\Delta p_0 \ = \ \Delta k  \frac{\gamma \hbar \sqrt{|\beta|}}{m_p} \ = \ \frac{\gamma \hbar \sqrt{|\beta|}}{m_p}\, .
\end{eqnarray}
With this we may write
\begin{eqnarray}
\sum_k \dots \ = \sum_k \Delta k \dots  = \ \sum_{p_0} \Delta p_0 \ \!\frac{m_p}{\gamma \hbar \sqrt{|\beta|}} \dots \, , 
\end{eqnarray}
where $p_0$ is discrete with values $-m_p/\sqrt{|\beta|} + k \gamma \hbar \sqrt{|\beta|}/m_p $.
This fact was used on the last line in (\ref{c.12.cc}).
It is important to bear in mind that the Cauchy principal value integral should be utilised in the $p_0$ integral in (\ref{c.12.cc}) in  order to 
see that the endpoint singularities  are integrable.


Note that in  the $|\beta| \rightarrow 0$ limit (\ref{c.12.cc}) reduces to
\begin{eqnarray}
&&\mbox{\hspace{-8.5mm}}\mathds{1}  \ = \ \int_{\mathds{R}} dx_0 \int_{\mathds{R}}  dp_0 \ \! \frac{1}{2\pi \hbar} \ \! |\psi, p_0,x_0\rangle\langle \psi, x_0, p_0|\, ,~~~
\end{eqnarray}
and hence
\begin{eqnarray}
\mu(p_0)|_{|\beta| \rightarrow 0}  \ = \ \frac{1}{2\pi \hbar}\, ,~~~~
\end{eqnarray}
which is the conventional value for (normalized) Schr\"{o}dinger wave packets. 

Should we have worked with states that are not normalized to unity (which is typical, e.g. for Glauber CSs~\cite{BJV}), we could assimilate the normalization factor to the measure, in which case we would obtained [cf. (\ref{37.ccf}) and (\ref{c.11.cff})]
\begin{eqnarray}
&&\mbox{\hspace{-12mm}}\mu(x_0,p_0)_{\rm{NN}} \ = \ \mu(p_0)_{\rm{NN}}  \ = \ \tilde{\mu}(y)_{\rm{NN}} \nonumber \\[2mm]
&&\mbox{\hspace{-6mm}}\propto \ \frac{\sqrt{a}}{2^a} \ \!\frac{\Gamma\left(a+1\right) \ \!\sum_{k=0}^{\infty} \delta(\pm y + a/2 -k)}{\Gamma(1 +\frac{a}{2} + y)\Gamma(1 +\frac{a}{2} - y)}\, ,
\end{eqnarray}
which in the $|\beta| \rightarrow 0$ limit reduces to
\begin{eqnarray}
{\mu}(p_0)_{\rm{NN}, |\beta| \rightarrow 0 } \ \propto \ \exp\left(-\frac{p_0^2}{\gamma \hbar}   \right)\, ,
\end{eqnarray}
as one would expect.


\begin{thebibliography}{99}
\smallskip

\bibitem{VenezGrossMende}
D.~Amati, M.~Ciafaloni and G.~Veneziano,
Phys.\ Lett.\ B {\bf 197}, 81 (1987).
\bibitem{Gross}
D.~J.~Gross and P.~F.~Mende,
Phys.\ Lett.\ B {\bf 197}, 129 (1987).
\bibitem{Ciafa}
D.~Amati, M.~Ciafaloni and G.~Veneziano,
Phys.\ Lett.\ B {\bf 216}, 41 (1989).
\bibitem{Koni}
K.~Konishi, G.~Paffuti and P.~Provero,
Phys.\ Lett.\ B {\bf 234}, 276 (1990).
%
\bibitem{MM}
M.~Maggiore,
Phys.\ Lett.\ B {\bf 319}, 83 (1993).
\bibitem{Kempf}
A.~Kempf, G.~Mangano and R.~B.~Mann,
  Phys.\ Rev.\ D {\bf 52}, 1108 (1995).
\bibitem{FS}
F.~Scardigli,
  Phys.\ Lett.\ B {\bf 452}, 39 (1999).
%
\bibitem{Adler2}
R.~J.~Adler and D.~I.~Santiago,
  Mod.\ Phys.\ Lett.\ A {\bf 14}, 1371 (1999).
%
\bibitem{Capoz}
S.~Capozziello, G.~Lambiase and G.~Scarpetta,
Int.\ J.\ Theor.\ Phys.\  {\bf 39}, 15 (2000).
%
\bibitem{MS1}
 J.~Magueijo and L.~Smolin, Phys. \ Rev. \ Lett. {\bf 88}, 190403 (2002).
%
\bibitem{MS2}
 J.~Magueijo and L.~Smolin, Phys. \ Rev. \ D {\bf 67}, 044017 (2003). 
%
\bibitem{Bojo}
M.~Bojowald and A.~Kempf,
  Phys.\ Rev.\ D {\bf 86}, 085017 (2012).
 \bibitem{Das}
S.~Das and E.~C.~Vagenas,
Phys. Rev. Lett. \textbf{101}, 221301 (2008).
%
  \bibitem{Brau}
 F.~Brau,
  J.\ Phys.\ A {\bf 32} 7691 (1999).
\bibitem{Pedram}
P.~Pedram, K.~Nozari and S.~H.~Taheri,
  JHEP {\bf 1103}, 093 (2011).
%
\bibitem{ScardCas}
F.~Scardigli and R.~Casadio,
   Eur.\ Phys.\ J.\ C {\bf 75}, 425 (2015).
 %
 \bibitem{Bosso}
  P.~Bosso, S.~Das and R.~B.~Mann,
Phys. Rev. D \textbf{96}, 066008 (2017).
%
\bibitem{QC}
F.~Scardigli, G.~Lambiase and E.~Vagenas,
  Phys.\ Lett.\ B {\bf 767}, 242 (2017).
%
\bibitem{Scardigli:2018jlm}
F.~Scardigli, M.~Blasone, G.~Luciano and R.~Casadio,
Eur. Phys. J. C \textbf{78}, 728 (2018).
%
\bibitem{Petroz}
G.~G.~Luciano and L.~Petruzziello,
Eur. Phys. J. C \textbf{79},  283 (2019).
%
\bibitem{DaRocha}
I.~Kuntz and R.~Da Rocha,
Eur. Phys. J. C \textbf{80}, 478 (2020).
%
\bibitem{Buoninf}
L.~Buoninfante, G.~Lambiase, G.~G.~Luciano and L.~Petruzziello,
Eur. Phys. J. C \textbf{80}, 853 (2020).
%
\bibitem{CGgrav}
L.~Buoninfante, G.~G.~Luciano and L.~Petruzziello,
Eur. Phys. J. C \textbf{79}, 663 (2019).
%
\bibitem{Chung:2019raj}
W.~S.~Chung and H.~Hassanabadi,
Eur. Phys. J. C \textbf{79}, 213 (2019).
%
\bibitem{fabian}
F.~Wagner,
Phys. Rev. D \textbf{104}, 126010 (2021).
%
\bibitem{LucBosFri}
P.~Bosso, M.~Fridman and G.~G.~Luciano,
Front. Astron. Space Sci. \textbf{9}, 932276 (2022).
%
\bibitem{Bruk}
 I.~Pikovski, M.~R.~Vanner, M.~Aspelmeyer, M.~S.~Kim and C.~Brukner,
Nature Phys. \textbf{8}, 393 (2012)
%
\bibitem{AliTest}
A.~F.~Ali, S.~Das and E.~C.~Vagenas,
Phys. Rev. D \textbf{84}, 044013 (2011).
%
\bibitem{Bawaj}
M.~Bawaj, C.~Biancofiore, M.~Bonaldi, F.~Bonfigli, A.~Borrielli, G.~Di Giuseppe, L.~Marconi, F.~Marino, R.~Natali and A.~Pontin, \textit{et al.}
Nature Commun. \textbf{6}, 7503 (2015).
%
\bibitem{Pendu}
P.~A.~Bushev, J.~Bourhill, M.~Goryachev, N.~Kukharchyk, E.~Ivanov, S.~Galliou, M.~E.~Tobar and S.~Danilishin,
Phys. Rev. D \textbf{100}, 066020 (2019).
 %
 \bibitem{GravBar}
F.~Marin, F.~Marino, M.~Bonaldi, M.~Cerdonio, L.~Conti, P.~Falferi, R.~Mezzena, A.~Ortolan, G.~A.~Prodi and L.~Taffarello, \textit{et al.}
Nature Phys. \textbf{9}, 71 (2013).
%
 \bibitem{BossoLigo}
P.~Bosso, S.~Das and R.~B.~Mann,
Phys. Lett. B \textbf{785},  498 (2018).
%
\bibitem{LucLuc}
G.G.~Luciano and L.~Petruzziello,
Eur. Phys. J. Plus \textbf{136}, 179 (2021). 
%
\bibitem{Primord}
G.~G.~Luciano,
Eur. Phys. J. C \textbf{81}, 1086 (2021).
%
\bibitem{JKS}
P.~Jizba, H.~Kleinert and F.~Scardigli, Phys.\ Rev.\ D {\bf 81}, 084030 (2010).
%
\bibitem{KLVY}
T. Kanazawa, G. Lambiase, G. Vilasi, and A. Yoshioka, Eur. Phys. J. C {\bf 79}, 95 (2019).
%
\bibitem{Ong}
Y.~C.~Ong,
JCAP \textbf{09}, 015 (2018).
%
\bibitem{illu}
L.~Petruzziello and F.~Illuminati,
Nature Commun. \textbf{12}, 4449 (2021).
%
\bibitem{illu2}
E.~Al-Nasrallah, S.~Das, F.~Illuminati, L.~Petruzziello and E.~C.~Vagenas,
Nucl. Phys. B \textbf{992}, 116246 (2023).
%
\bibitem{Maggiore:93}
M.~Maggiore, Phys. Lett. B \textbf{304}, 65 (1993).
  %
\bibitem{Schrodinger} E.~Schr\"{o}dinger, {\em Collected Papers on Wave Mechanics}, (Blackie \& Son, London, 1928), p.41.
%
\bibitem{Robertson} H.P.~Robertson, Phys.\ Rev. \ {\bf 34}, 163 (1929).
%
\bibitem{Muk} N.~Mukhopadhyay, {\em Probability and Statistical Inference}, (Marcel Dekker, Inc., New York, 2000).
%
\bibitem{Bernardo} 
 R.C.S.~Bernardo and J.P.H.~Esguerra, Annals \ Phys. {\bf 391}, 293 (2018).
 %
 \bibitem{Panella}
A.~M.~Frassino and O.~Panella,
  Phys.\ Rev.\ D {\bf 85}, 045030 (2012).
  %
\bibitem{Ijmpd}
M.~Blasone, G.~Lambiase, G.G.~Luciano, L.~Petruzziello and F.~Scardigli,
  Int.\ J.\ Mod.\ Phys.\ D {\bf 29}, 2050011 (2020).
  %
\bibitem{bpw}
P.~Bosso, L.~Petruzziello and F.~Wagner,
Phys. Lett. B \textbf{834}, 137415 (2022)
%
\bibitem{Husain}
V. Husain, D. Kothawala and S.S. Seahra Phys. Rev. D
\textbf{87}, 025014 (2013).
%
\bibitem{BossoLuciano}
P.~Bosso and G.~G.~Luciano,
Eur. Phys. J. C \textbf{81}, 982 (2021).
%
\bibitem{JizSc}
P. Jizba and F. Scardigli, Phys. Rev. D \textbf{86}, 025029
(2012)
%
\bibitem{Giardino}
S.~Giardino and V.~Salzano,
Eur. Phys. J. C \textbf{81}, 110 (2021).
%
\bibitem{DFLV}
S. Das, M. Fridman, G. Lambiase, and E.C. Vagenas, Phys. Lett. B {\bf 824}, 136841 (2022). 
%
\bibitem{Tawfik}
A. Tawfik and A. Diab, Int. J. Mod. Phys. D \textbf{23},
1430025 (2014).
%
\bibitem{Braidotti}
M.C. Braidotti, Z.H. Musslimani and C. Conti, Physica
D \textbf{338}, 34 (2017).
%
\bibitem{iorio}
A.~Iorio, B.~Iveti\'c, S.~Mignemi and P.~Pais,
Phys. Rev. D \textbf{106}, 116011 (2022).
%
\bibitem{Kiefer} C.~Kiefer, {\em Quantum Gravity}, (Oxford Science Publications, Oxford, 2012).
%
\bibitem{Kiefer2} C.~Kiefer, {\em The semiclassical approximation to quantum gravity}. In: Ehlers J., Friedrich H. (eds) {\em Canonical Gravity: From Classical to Quantum}, Lecture Notes in Physics, vol 434. (Springer, Berlin, 1994).
%
\bibitem{Dutra}
S.M. Dutra, J. Mod. Optic. \textbf{45}, 759 (1998).
%
\bibitem{Matacz}
A.L. Matacz, Phys. Rev. D \textbf{49}, 788 (1994).
%
\bibitem{Mandel}
L. Mandel and E. Wolf, {\em Optical Coherence and Quantum
Optics} (Cambridge University Press, Cambridge, 1995).
%
%
\bibitem{Chang}
L.N. Chang, D. Minic, N. Okaruma, T. Takeuchi, Phys. Rev. D
\textbf{65}, 125028 (2002).
%
%
%
%
%
%
\bibitem{Shababi}
H. Shababi, K. Ourabah, Eur. Phys. J. Plus \textbf{135}, 697 (2020).
%
\bibitem{LucianoTs}
G.G.~Luciano,
Eur. Phys. J. C \textbf{81}, 672 (2021).
%
\bibitem{Nieto:78} M.M.~Nieto and L.M.~Simmons, Jr., Phys. \ Rev. \ Lett. \ {\bf 41}, 207 (1978).
%
\bibitem{Tsallis1}
C.~Tsallis, J. Stat. Phys. {\bf 52}, 479 (1988).
  %
\bibitem{Tsallis2}
C. Tsallis, \emph{Introduction to Non-Extensive Statistical Mechanics:
Approaching a Complex World}, (Springer, Berlin, 2009).
%
\bibitem{Abe1}
S.~Abe, S.~Mart\'{i}nez, F.~Pennini and A.~Plastino, Phys. Lett. A {\bf 281}, 126 (2001).
%
\bibitem{jizba-korbel:19}
P.~Jizba and J.~Korbel, 
Phys. Rev. Lett. {\bf 122}, 120601 (2019). 
%
%
\bibitem{JKZ} P.~Jizba, J.~Korbel and V.~Zatloukal, Phys. \ Rev. \ E {\bf 95}, 022103 (2017).
%
\bibitem{Caratheodory} C. Caratheodory, Math. Ann. {\bf 67} 355 (1909).
%
\bibitem{CaratheodoryII} H.A. Buchdahl, Am. J. Phys. {\bf 17} 212 (1949).
%
\bibitem{Tsallis3}
C.~Tsallis and L.~J.~L.~Cirto, Eur. Phys. J. C {\bf 73}, 2487 (2013).
%
\bibitem{Verlinde}
E.P.~Verlinde,
JHEP \textbf{04}, 029 (2011).
%
\bibitem{mannheim}
P.D.~Mannheim and J.G.~O'Brien,
Phys. Rev. Lett. \textbf{106}, 121101 (2011).
%
\bibitem{mannheim_a} P.D.~Mannheim and J.G.~O'Brien,
Phys. Rev. D \textbf{85}, 124020 (2012).
%
\bibitem{mannheim_b}
J.G.~O'Brien and P.D.~Mannheim,
Mon. Not. Roy. Astron. Soc. \textbf{421}, 1273 (2012).
\bibitem{mannheim_c}
J.G.~O'Brien, T.L.~Chiarelli and P.D.~Mannheim,
Phys. Lett. B \textbf{782}, 433-439 (2018).	
%
\bibitem{Kazanas} J.~Sultana, D.~Kazanas and J.L.~Said, Phys. Rev. D \textbf{86}, 084008 (2012).
%
\bibitem{JLLP}
P.~Jizba, G.~Lambiase, G.G.~Luciano and L.~Petruzziello, Phys. Rev. D \textbf{105}, L121501 (2022).
%
%
\bibitem{Thistleton}
W.~Thistleton, J.A.~Marsh, K.~Nelson and C.~Tsallis,
IEEE Trans. Inf. Theory {\bf 53}, 4805 (2007).
%
%
\bibitem{BJV} M.~Blasone, P.~Jizba and G.~Vitiello, {\em Quantum Field Theory and its Macroscopic Manifestations}, (World Scientific \& ICP, London, 2010).
%
\bibitem{perelomov}
A.~Perelomov, {\em Generalized Coherent States and Their Applications}, (Springer-Verlag, Berlin, 1986).
%
\bibitem{BG}
A.O.~Barut and L.~Girardello, Commun. Math. Phys. {\bf 21}, 41 (1971). 
%
\bibitem{nieto}
M.M.~Nieto and L.M.~Simmons, Jr., Phys. Rev. Lett. {\bf 41}, 207 (1978).
%
%
\bibitem{KlauderIIa}
J.R.~Klauder and B.-S.~Skagerstam (editors), {\em Coherent States: Applications in Physics and Mathematical Physics}, (Singapore: World Scientific, 1985).
%
%
\bibitem{Ramanujan} S.~Ramanujan,  Messenger of Mathematics, {\bf 44}, 10 (1915).
%
%
\bibitem{zurek1} J.P.~Paz and W.H.~Zurek, Phys. Rev. Lett. {\bf 82}, 5181 (1999).
%
\bibitem{zurek2} A.~Venugopalan, Phys. Rev. A {\bf 61}, 012102 (1999).

\bibitem{zurek3} D.A.R.~Dalvit, J.~Dziarmaga, and W.H.~Zurek, Phys. Rev. A {\bf 72}, 062101 (2005). 
%
\bibitem{zurek4} J.P.~Paz and W.H.~Zurek, in Coherent Atomic Matter Waves. Les Houches-Ecole d'Ete de Physique Theorique, edited by R.~Kaiser, C.~Westbrook, and F.~David (Springer, Berlin, 2001), Vol. 72. 
%
%
\bibitem{Huang}
K.~Huang, \emph{Statistical Mechanics}, (John Wiley \& Sons, New York, 1987).
%
\bibitem{Renyi}
A.~R\'{e}nyi, \emph{Selected Papers of Alfred R\'{e}nyi, vol. 2}, (Akademia Kiado, Budapest, 1976).
%
\bibitem{JD:16} P.~Jizba, Y.~Ma, A.~Hayes and J.A.~Dunningham, Phys.~Rev.~E \textbf{93},  060104(R) (2016).
%
\bibitem{JDJ} P.~Jizba, J.A.~Dunningham and J.~Joo, Annals of Physics {\bf 355},  87 (2015). 
%
\bibitem{JDP} P.~Jizba, J.A.~Dunningham and M.~Prok\v{s},
 Entropy {\bf 23}, 334 (2021).
%
\bibitem{Shannon} C.E.~Shannon, \emph{ A mathematical theory of communication},
Bell Syst. Tech. J. \textbf{27},   379; 623 (1948).
%
\bibitem{Beckner1975} W.~Beckner, Ann. of Math. \textbf{102},  159 (1975).
%
\bibitem{Babenko1962} K.I.~Babenko, Amer. Math. Soc. Transl. Ser. 2 \textbf{44}, 115 (1962).
%
%
\bibitem{Hirschman}
I.I.~Hirschman, Jr.,  Am. J. Math. {\bf 79}, 152 (1957).
%
\bibitem{BB}
I.~Bialynicki--Birula and J.~Mycielski, J. Commun. Math. Phys. {\bf 44}, 129 (1975).
%
%
\bibitem{angmom}
P.~Bosso and S.~Das,
Annals Phys. \textbf{383}, 416 (2017).
%
\bibitem{angmom2}
P.~Bosso, L.~Petruzziello, F.~Wagner and F.~Illuminati,
Commun. Phys. \textbf{6}, 114 (2023)
%
\bibitem{Hill} T.L.~Hill, {\em Thermodynamics of Small Systems}, (Dover, New
York, 1994).
%
\bibitem{MK2} P.D.~Mannheim and D.~Kazanas, Astrophys. J. {\bf 342}, 635
(1989).
%
%
%
%

%
%
\bibitem{Burges} C.P.~Burgess, R.~Holman and D. Hoover, Phys. Rev. D {\bf 77}, 063534 (2008).
%
\bibitem{KieferII} C.~Kiefer, I.~Lohmar, D.~Polarski and A.A.~Starobinsky,
Classical Quantum Gravity {\bf 24}, 1699 (2007).
%
%
\bibitem{Du:04} J.~Du, Physica (Amsterdam) {\bf 335A}, 107 (2004).
%
\bibitem{Abe4} S.~Abe, Phys. Lett. A {\bf 263}, 424 (1999); {\bf 267}, 456(E) (2000).
%
\bibitem{Bar:16}
J.D.~Barrow and A.~Paliathanasis, Phys. Rev. D {\bf 94}, 083518 (2016).
%
\bibitem{Bar:18}
J.D.~Barrow and A.~Paliathanasis, Gen. Relativ. Gravit. {\bf 50}, 82 (2018).
%
\bibitem{VF} A.~Vilenkin and L.H.~Ford, Phys. Rev. D {\bf 26}, 1231 (1982).
%
%
\bibitem{Rovelli}
C.~Rovelli,
Living Rev. Rel. \textbf{1}, 1 (1998).
%
%
\bibitem{Rovelli2}
C. Rovelli, L. Smolin, 
Nucl. Phys. B {\bf442} 593 (1995). 
%
\bibitem{Immirzi}
A.~Ashtekar, J.~C.~Baez and K.~Krasnov,
Adv. Theor. Math. Phys. \textbf{4}, 1 (2000).
%
\bibitem{QNM}
O.~Dreyer,
Phys. Rev. Lett. \textbf{90}, 081301 (2003). 
%
\bibitem{ReW}
T. Regge, J.A. Wheeler, Phys. Rev. {\bf108}, 1063  (1957).
%
\bibitem{AbreuQN}
E.M.C.~Abreu, J.A.~Neto, E.M.~Barboza and B.B.~Soares,
Phys. Lett. B \textbf{798}, 135011 (2019).
%
\bibitem{Damp}
H.-P. Nollert, Phys. Rev. D {\bf 47} 5253 (1993).
%
\bibitem{Hod}
S. Hod, Phys. Rev. Lett. {\bf 81} 4293 (1998). 
%
\bibitem{coraddu}
M.~Coraddu and S.~Mignemi,
EPL \textbf{91}, 51002 (2010).
%
\bibitem{curved1}
G.~Amelino-Camelia,
Phys. Lett. B \textbf{510}, 255 (2001).
\bibitem{curved2}
J.~Kowalski-Glikman,
Phys. Lett. B \textbf{547}, 291 (2002).
\bibitem{curved3}
D.~Raetzel, S.~Rivera and F.~P.~Schuller,
Phys. Rev. D \textbf{83}, 044047 (2011).
\bibitem{curved4}
G.~Amelino-Camelia, L.~Freidel, J.~Kowalski-Glikman and L.~Smolin,
Phys. Rev. D \textbf{84}, 084010 (2011).
\bibitem{curved5}
L.~Smolin,
Gen. Rel. Grav. \textbf{43}, 3671 (2011).
\bibitem{curved6}
H.~J.~Matschull and M.~Welling,
Class. Quant. Grav. \textbf{15}, 2981 (1998).
\bibitem{curved7}
G.~Amelino-Camelia, M.~Arzano, S.~Bianco and R.~J.~Buonocore,
Class. Quant. Grav. \textbf{30}, 065012 (2013).
%
\bibitem{fab1}
F.~Wagner,
Phys. Rev. D \textbf{104}, 126010 (2021).
%
\bibitem{fab2}
F.~Wagner,
Eur. Phys. J. C \textbf{83},  154 (2023).
%
\bibitem{sb1}
G.~Amelino-Camelia, L.~Smolin and A.~Starodubtsev,
Class. Quant. Grav. \textbf{21}, 3095 (2004).
%
\bibitem{sb2}
S.~Hossenfelder,
Phys. Rev. D \textbf{75}, 105005 (2007).
%
\bibitem{sb3}
G.~Amelino-Camelia, L.~Freidel, J.~Kowalski-Glikman and L.~Smolin,
Phys. Rev. D \textbf{84}, 087702 (2011).
%
\bibitem{sb4}
S.~Hossenfelder,
SIGMA \textbf{10}, 074 (2014).
%
\bibitem{sb5}
G.~Amelino-Camelia,
Entropy \textbf{19}, 400 (2017).
%
\bibitem{sb6}
G.~Amelino-Camelia, V.~Astuti, M.~Palmisano and M.~Ronco,
Int. J. Mod. Phys. D \textbf{30}, 2150046 (2021).
%
\bibitem{camel}
G.~Amelino-Camelia,
Phys. Rev. Lett. \textbf{111}, 101301 (2013).
%
\bibitem{Kaniadakis1}
G.~Kaniadakis,
Phys. Rev. E \textbf{66}, 056125 (2002). 
%
\bibitem{Kaniadakis2}
G. Kaniadakis, Physica A {\bf 296}, 405 (2001).
%
\bibitem{Kaniadakis3}
G.~G.~Luciano,
Eur. Phys. J. C \textbf{82}, 314 (2022).
%
\bibitem{RevLucia}
G.G. Luciano, Entropy {\bf 24}, 1712 (2022). 
%
%
%
\bibitem{distributions} N.L.~Johnson, S.~Kotz and N.~Balakrishnan, \emph{Continuous Univariate Distributions, Volume 2}, John Wiley \& Sons, Inc.
New York, 1995.
%
\bibitem{Ober} F.~Oberhettinger, \emph{Fourier Transiorms of Distributions and Their Inverses, A Collection of Tables},
(Academic Press, New York, 1973).
%
%
%
\bibitem{Laplace-transf}
e.g., L.~Debnath and J.~Thomas~Jr., Z. Angew. Math. und Mech. {\bf 66}, 559 (1976).
%
\bibitem{ramanujan_b}
S.~Ramanujan, Quart. J. Math. {\bf 48A}, 294 (1920).
%
\bibitem{vladimirov}
V.S.~Vladimirov, Equations of Mathematical Physics,  (Marcel Dekker, Inc., New Yourk, 1971).


\end{thebibliography}
\end{document}